\documentclass[journal]{IEEEtran}

\usepackage{graphicx}
\usepackage{amssymb}
\usepackage{amsmath}
\usepackage{booktabs}
\usepackage{mathtools}
\usepackage{makecell}
\usepackage{xcolor}
\usepackage[space]{cite}
\usepackage[T1]{fontenc}

\usepackage[colorinlistoftodos]{todonotes}

\graphicspath{ {images/} }

\begin{document}

\title{Sound Event Detection and Time-Frequency Segmentation from Weakly Labelled Data}

\author{Qiuqiang Kong*, Yong Xu* \thanks{* The first two authors contributed equally to this work.}, Iwona Sobieraj, Wenwu Wang, Mark D. Plumbley \IEEEmembership{Fellow,~IEEE}
}

\markboth{}%
{Shell \MakeLowercase{\textit{et al.}}: Bare Demo of IEEEtran.cls for IEEE Journals}

\maketitle

\begin{abstract}
 Sound event detection (SED) aims to detect when and recognize what sound events happen in an audio clip. Many supervised SED algorithms rely on strongly labelled data which contains the onset and offset annotations of sound events. However, many audio tagging datasets are weakly labelled, that is, only the presence of the sound events is known, without knowing their onset and offset annotations. In this paper, we propose a time-frequency (T-F) segmentation framework trained on weakly labelled data to tackle the sound event detection and separation problem. In training, a segmentation mapping is applied on a T-F representation, such as log mel spectrogram of an audio clip to obtain T-F segmentation masks of sound events. The T-F segmentation masks can be used for separating the sound events from the background scenes in the time-frequency domain. Then a classification mapping is applied on the T-F segmentation masks to estimate the presence probabilities of the sound events. We model the segmentation mapping using a convolutional neural network and the classification mapping using a global weighted rank pooling (GWRP). In SED, predicted onset and offset times can be obtained from the T-F segmentation masks. As a byproduct, separated waveforms of sound events can be obtained from the T-F segmentation masks. We remixed the DCASE 2018 Task 1 acoustic scene data with the DCASE 2018 Task 2 sound events data. When mixing under 0 dB, the proposed method achieved F1 scores of 0.534, 0.398 and 0.167 in audio tagging, frame-wise SED and event-wise SED, outperforming the fully connected deep neural network baseline of 0.331, 0.237 and 0.120, respectively. In T-F segmentation, we achieved an F1 score of 0.218, where previous methods were not able to do T-F segmentation.

\end{abstract}


\IEEEpeerreviewmaketitle

\section{Introduction} \label{section:introduction}
Sound event detection (SED) aims to detect what sound events happen in an audio recording and when they occur. SED has many applications in everyday life. For example, SED can be used to monitor ``baby cry'' sound at home \cite{saraswathy2012automatic}, and to detect ``typing keyboard'', ``door slamming'', ``ringing of phones'', ``smoke alarms'' and ``sirens'' in the office \cite{harma2005automatic, ellis2001detecting}. For public security, SED can be used to detect ``gunshot'' and ``scream'' sounds \cite{valenzise2007scream}. Not only is SED complementary to video or image based event detection \cite{xu2015discriminative, girshick2014rich, borji2015salient} but also has many advantages over the two modalities. First, sound does not require illumination, so can be used in dark environments. Second, sound can penetrate or move around some obstacles, while objects in video and image are often occluded. Third, some abnormal events such as fire alarms are audio only, so can only be detected by sound. Furthermore, storing and processing sound often consumes less computation resources than video \cite{abu2016youtube}, and as a result, longer sound sequences can be stored in a device and faster processing can be obtained using equal computation resources. 
\begin{figure}[t]
  \centering
  \centerline{\includegraphics[width=\columnwidth]{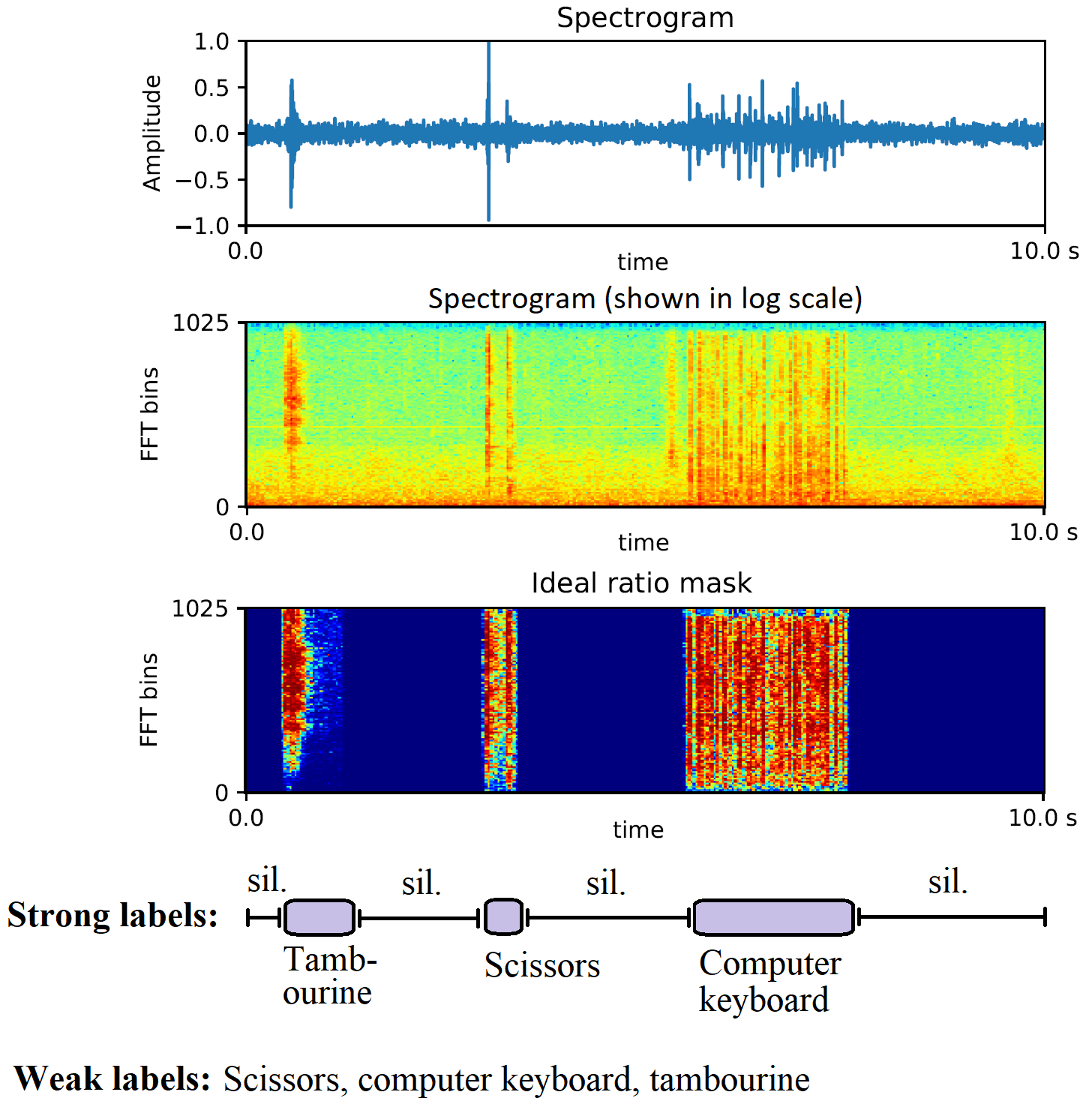}}
  \caption{From top to bottom: Waveform of an audio clip containing three sound events: ``Tambourine'', ``scissors'' and ``computer keyboard''; Log mel spectrogram of the audio clip; Ideal ratio mask (IRM) \cite{narayanan2013ideal} of sound events. Strongly labelled onset and offset annotations of sound events; Weak labels. ``Silence'' is the abbreviated as ``sil.''. The signal-to-noise ratio of this audio clip is 0 dB.}
  \label{fig:waveform}
\end{figure}
\begin{figure*}[t]
  \centering
  \centerline{\includegraphics[width=\textwidth]{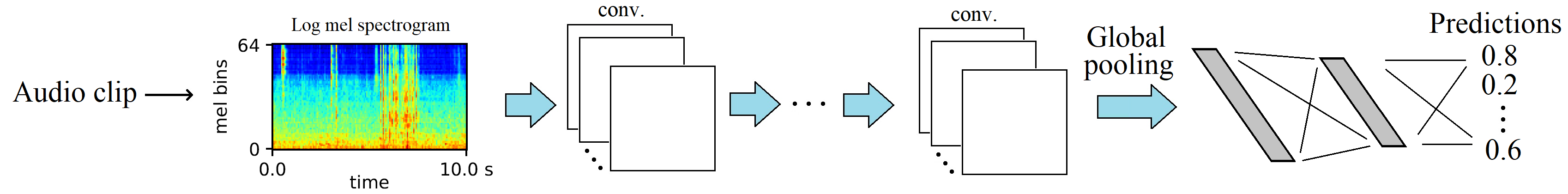}}
  \caption{Audio tagging with convolutional neural network. Input log mel spectrogram is presented to a convolutional neural network including convolutional layers, a global pooling layer and fully connected layers to predict the presence probabilities of audio tags. }
  \label{fig:cnn}
\end{figure*}
Many SED algorithms rely on \textit{strongly labelled data} \cite{stowell2015detection, parascandolo2016recurrent, mesaros2016tut} where the onset and offset times of sound events have been annotated. The segments between the onset and offset labels are used as target events for training, while those outside the onset and offset annotations are used as non-target events \cite{parascandolo2016recurrent, mesaros2016tut}. However, collecting strongly labelled data is time consuming because annotating the onset and offset times of sound events takes more time than annotating audio clips for classification, so the sizes of strongly labelled datasets are often limited to minutes or a few hours \cite{mesaros2016tut, giannoulis2013detection}. At the same time there are large amounts of \textit{weakly labelled data} (WLD) available, where only the presence of the sound events is labelled, without any onset and offset annotations \cite{kumar2016audio, 2017_Adavanne_Sound} or the sequence of the sound events. Fig. \ref{fig:waveform} shows the waveform of an audio clip containing three non-overlapping sound events, the log mel spectrogram of the audio clip, the ideal ratio mask (IRM) \cite{narayanan2013ideal} of the sound events, the strongly labelled onset and offset annotations and the weak labels. In this paper we will focus on non-overlapping sound events as a starting point. 

In the real world, sound events usually happen in real scenes such as a metro station or an urban park. State-of-the-art SED algorithms only detect the onset and the offset of sound events in the time domain but do not separate them from background in the T-F domain. The separation of sound events in the T-F domain can be useful for enhancing and recognizing sound events in audio scenes under low signal-to-noise ratio (SNR). In this paper, we propose a T-F segmentation and sound event detection framework trained using weakly labelled data.
This is done by learning \textit{T-F segmentation masks} implicitly in training with only the clip-level audio tags. It means that T-F masks are not known even for the training set: they are predicted as intermediate results. 
T-F segmentation masks are equivalent to the ideal ratio masks (IRM) \cite{narayanan2013ideal}. An IRM is the ratio of the spectrogram of a sound event to the spectrogram of the mixed audio. T-F segmentation masks can be used for SED and sound event separation. In training, a \textit{segmentation mapping} is applied to the T-F representation such as log mel spectrogram of an audio clip to obtain T-F segmentation masks for sound events. Then a \textit{classification mapping} is applied to the T-F segmentation masks to output the presence probabilities of sound events. In T-F segmentation, with a T-F representation of an audio clip as input, the trained segmentation mapping is used to obtain the T-F segmentation masks. In SED, onset and offset times can be obtained from the T-F segmentation masks. As a byproduct, separated waveforms of sound events can be obtained from the T-F segmentation masks. This work is an extension of the joint separation-classification model for SED of weakly labelled data \cite{kong2017joint_separation}. 

The paper is organized as follows. Section \ref{section:related_works} introduces previous work in SED with WLD. Section \ref{section:wld} describes the proposed T-F segmentation, sound event detection and separation framework. Section \ref{section:implementation} describes the implementation details of the proposed framework. Section \ref{section:experiments} shows experimental results. Section \ref{section:conclusion} concludes and forecasts future work.  

\section{Weakly supervised sound event detection} \label{section:related_works}
Compared to the conventional SED task, where strongly labelled onset and offset annotations for the training set are given, the weakly supervised SED task contains only clip-level labels. That is, only the presence of sound events is known in an audio clip, without knowing the temporal locations of the events. Several approaches for weakly supervised SED have recently been proposed, including multiple instance learning and convolutional neural networks. 

\subsection{Multi-instance learning method} \label{subsection:mil}
One solution to the WLD problem is based on multiple instance learning (MIL) \cite{maron1998framework, kumar2016audio}. MIL was first proposed in 1997 for drug activity detection \cite{dietterich1997solving}. In MIL for SED, an audio clip is labelled positive for a specified sound event if that sound event occurs at least one time in the audio clip, and labelled negative if that sound event does not occur in the audio clip. For strongly labelled data, the dataset consists of training pairs $ \{ x, y \} $ where $ x $ is the feature of a frame in an audio clip and $ y \in \{ 0, 1 \}^{K} $ is the strong label of the frame, where $ K $ denotes the number of sound classes. For weakly labelled data, features of all frames in an audio clip constitute a bag $ B = \{ x_{t}\}_{t=1}^{T} $ where $ T $ is the number of frames in the audio clip. Multiple instance assumption states that the weak labels of a bag are $y= \max\limits_{t}\{y_t\}_{t=1}^{T} $, where $ y_{t} $ is the strong label of the feature $ x_{t}$. The weakly labelled data consists of the training pairs $ \{ B, y \} $. 

The problem of SED from WLD now can be cast as learning a classifier to predict the labels of the frames $ \{ y_{t} \}_{t=1}^{T} $ of a bag $ B = \{ x_{t}\}_{t=1}^{T} $. For the general WLD problem, an MIL framework based on a neural network was proposed in \cite{zhou2002neural, kumar2016audio}. In \cite{andrews2003support, kumar2016audio} a support vector machine (SVM) was used to solve MIL as a maximum margin problem. A negative mining method was proposed in \cite{siva2012defence} that selects negative examples according to intra-class variance criterion. A concept ranking according to negative exemplars (CRANE) algorithm was proposed in \cite{tang2013discriminative}. However, an MIL method tends to underestimate the number of positive instances in an audio clip \cite{kolesnikov2016seed}. Furthermore, the MIL method cannot predict the T-F segmentations from the WLD \cite{kumar2016audio}. 
\begin{figure*}[t]
  \centering
  \centerline{\includegraphics[width=\textwidth]{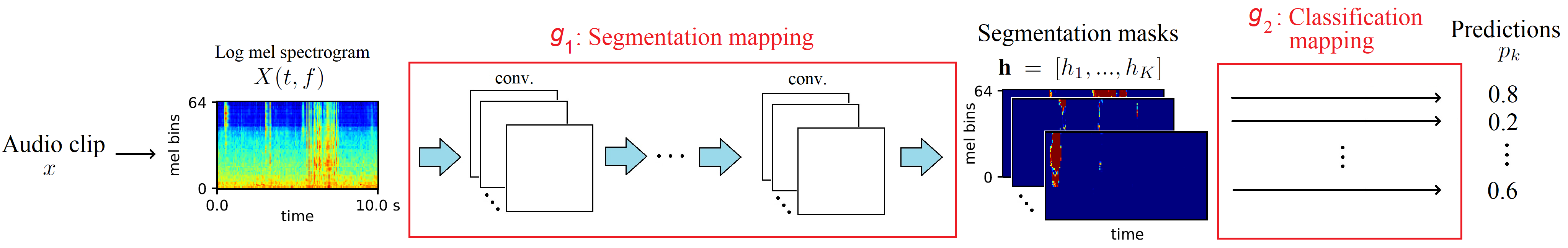}}
  \caption{Training stage using weakly labelled data. A segmentation mapping $ g_{1} $ maps from an input T-F representation to the segmentation masks. A classification mapping $ g_{2} $ maps each segmentation mask to the presence probabilities of the corresponding audio tag. }
  \label{fig:train_stage}
\end{figure*}
\subsection{Convolutional neural networks for audio tagging and weakly supervised sound event detection} \label{subsection:cnn}
Convolutional neural networks (CNNs) have been successfully used in many areas including image classification \cite{krizhevsky2012imagenet}, object detection \cite{girshick2014rich}, image segmentation \cite{long2015fully}, speech recognition \cite{dahl2012context, abdel2014convolutional} and audio classification \cite{choi2016automatic}. In this section we briefly introduce previous work using convolutional neural network for audio tagging \cite{choi2016automatic} and weakly supervised SED. 

Audio tagging \cite{choi2016automatic, foster2015chime, mesaros2016tut} aims to predict the presence of sound events in an audio clip. In \cite{zeiler2014visualizing}, a mel spectrogram of an audio clip is presented to a CNN, where the filters of each convolutional layer capture local patterns of a spectrogram. After a global pooling layer such as global max pooling \cite{choi2016automatic}, global average pooling \cite{lin2013network}, global weighted rank pooling \cite{kolesnikov2016seed}, global attention pooling \cite{kong2017audio, mcfee2018adaptive} or other poolings \cite{2017_Chou_FrameCNN, su2017weakly}, fully connected layers are applied to predict the presence probabilities of audio classes. Fig. \ref{fig:cnn} shows the framework of audio tagging with convolutional neural network. However, this CNN only predicts the presence probabilities of a sound events in an audio clip, but not the onset and offset times of the sound events. 

In \cite{tseng2017multiple, kumar2017deep}, a time-distributed CNN with a global max-pooling strategy was proposed to approximate the MIL method to predict the temporal locations of each event. However, the global max-pooling will encourage the model to attend to the most dominant T-F unit contributing to the presence of the sound event and ignore all of other T-F units. That is, the happening time of the sound events is underestimated. A method  for localizing the sound events in an audio clip by splitting the input into several segments based on the CNNs was presented in \cite{lee2017ensemble}. It splits an audio clip into several segments with the assumption that parts of the segments correspond to the clip-level labels. This assumption may be unreasonable due to the fact that some sound events may only occur at certain frames. Recently, an attention-based global pooling strategy using CNNs was proposed to predict the temporal locations \cite{xu2017large} for SED using WLD. However, attention-based global pooling can only predict the time domain segmentation, but not the T-F segmentation which will be firstly addressed in this paper.

\section{Time-frequency segmentation, sound event detection and separation from weakly labelled data} \label{section:wld}
In this section, we present a T-F segmentation, sound event detection and separation framework trained on weakly labelled audio data. Unlike the CNN method for audio tagging, we design a CNN to learn T-F segmentation masks of sound events from the weakly labelled data. 

\subsection{Training from weakly labelled data} 
We use only weakly labelled audio data to train the proposed model. The training stage is shown in Fig. \ref{fig:train_stage}. To begin with, the waveform of an audio clip $ x $ is converted to an input time-frequency (T-F) representation $ X(t, f) $, for example, spectrogram or log mel spectrogram. To simplify the notation, we abbreviate $ X(t, f) $ as $ X $. 

The first part of the training stage is a \textit{segmentation mapping} $ g_{1}: X \mapsto \textbf{h} $ which maps the input T-F representation to the T-F segmentation masks $ \textbf{h} = [h_{1}, ..., h_{K}] $, where $ K $ is the number of T-F segmentation masks and is equal to the number of sound events. Symbol $ h_{k} $ is the abbreviation of $ h_{k}(t, f) $ which is the T-F segmentation mask of the $ k $-th event. Ideally, each T-F segmentation mask $ h_{k} $ is an ideal ratio mask \cite{narayanan2013ideal} of the $ k $-th sound event. 

The second part of the training stage is a \textit{classification mapping} $ g_{2}: h_{k} \mapsto p_{k}, k=1, ..., K $ where $ g_{2} $ maps each T-F segmentation mask to the presence probability of the $ k $-th event, denoted as $ p_{k} $. Then the binary crossentropy between the predictions $ p_{k}, k=1, ..., K $ and the targets $ y_{k}, k=1, ..., K $ is calculated as the loss function:
\begin{equation} \label{eq:binary_crossentropy}
\begin{split}
l \left ( p_{k}, y_{k} \right ) & = -\sum_{k=1}^{K}y_{k} \log p_{k} \\
& = -\sum_{k=1}^{K}y_{k} \log g_{2}(g_{1}(X)_{k}), 
\end{split}
\end{equation}
\noindent where $ y_{k} \in \{0, 1\}, k=1, ..., K $ is the binary representation of the weak labels. Both $ g_{1} $ and $ g_{2} $ can be modeled by neural networks. The parameters of $ g_{1} $ and $ g_{2} $ can be trained end-to-end from the input T-F representation to the weak labels of an audio clip. 
\begin{figure}[t!]
  \centering
  \centerline{\includegraphics[width=\columnwidth]{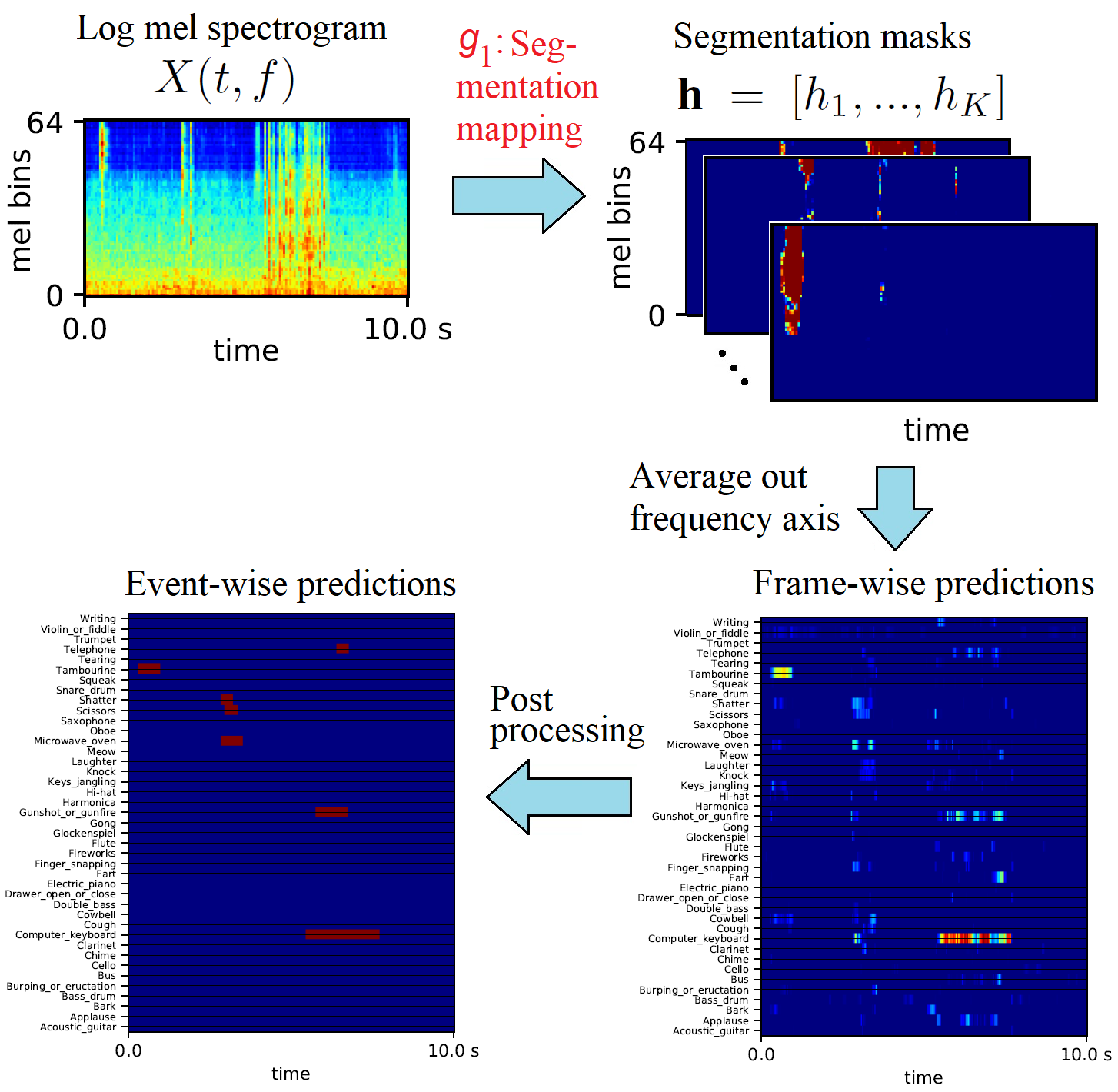}}
  \caption{Inference stage. An input T-F representation is presented to the segmentation mapping $ g_{1} $ to obtain the T-F segmentation masks. By averaging out the frequency axis of the T-F segmentation masks and post processing, event-wise predictions of sound events can be obtained. }
  \label{fig:fig_sed}
\end{figure}
\begin{figure*}[t!]
  \centering
  \centerline{\includegraphics[width=0.9\textwidth]{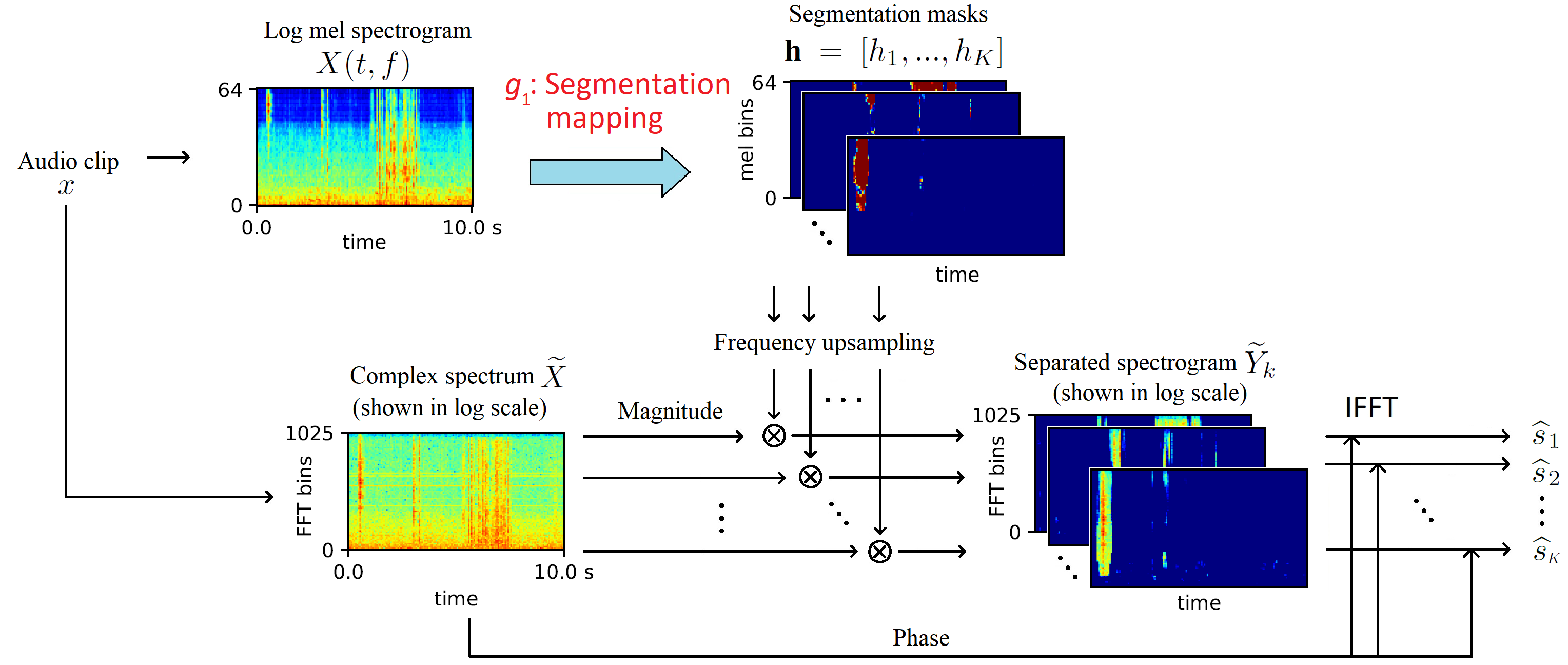}}
  \caption{Sound event separation stage. An input T-F representation is presented to the segmentation mapping $ g_{1} $ to obtain the T-F segmentation masks. The upsampled segmentation masks are multiplied with the magnitude spectrum of the input audio to obtain the segmented spectrogram of each sound event. Separated sound events are obtained by applying an inverse Fourier transform to the segmented spectrogram. }
  \label{fig:fig_ss}
\end{figure*}
\begin{figure}[t!]
  \centering
  \centerline{\includegraphics[width=\columnwidth]{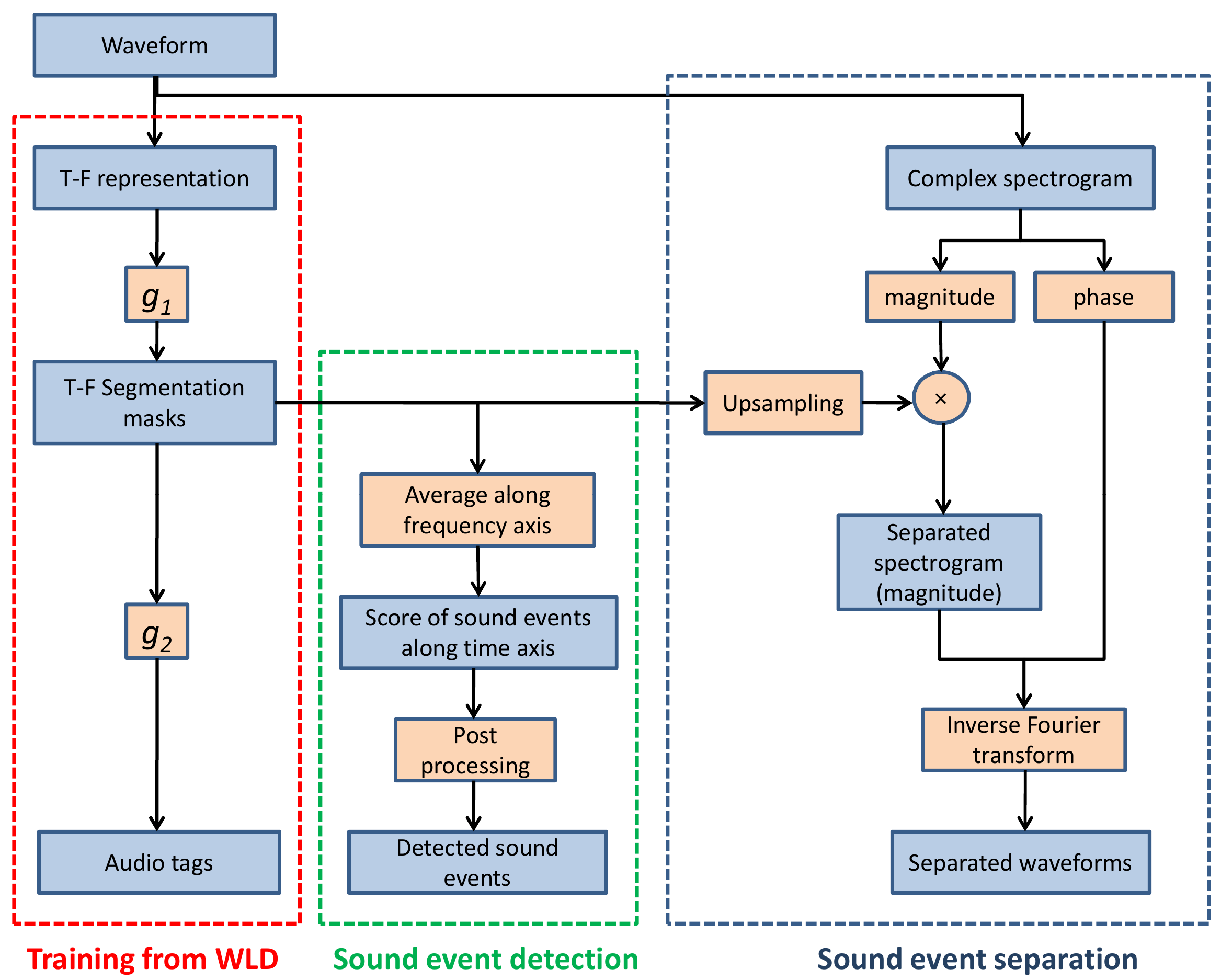}}
  \caption{Framework of T-F segmentation, sound event detection and sound event separation from WLD. From left to right: Training from WLD; Sound event detection; Sound event separation. }
  \label{fig:framework}
\end{figure}

\subsection{Time-frequency segmentation}
In inference step, the input T-F representation of an audio clip is presented to the segmentation mapping $ g_{1} $ to obtain the T-F segmentation masks $ h_{k}, k=1, ..., K $. The T-F segmentation masks indicate which T-F units in the T-F representation contribute to the presence of the sound events (top right of Fig. \ref{fig:fig_sed}). The learned T-F segmentation masks are affected by the classification mapping $ g_{2} $ and will be discussed in Section \ref{section:implementation}. 

\subsection{Sound event detection}\label{section:framework}
As T-F segmentation masks $ h_{k}, k=1, ..., K $ contain the information about where sound events happen in the T-F domain, the simplest way to obtain the sound event detection score $ v_{k}(t) $ in the time domain is to average out the frequency axis of the T-F segmentation masks (bottom right of Fig. \ref{fig:fig_sed}):
\begin{equation} \label{eq:sed}
v_{k}(t)=\frac{1}{F}\sum_{f=1}^{F} h_{k}(t, f),
\end{equation}
\noindent where $ F $ is the number of frequency bins of the segmentation mask $ h_{k} $. Then $ v_{k}(t) $ is the score of the frame-wise prediction of the sound events. We describe how to convert the frame-wise scores to event-wise sound events in Section \ref{section:implementation_sed}. 
\begin{figure*}[t]
	\centering
	\includegraphics[width=\textwidth]{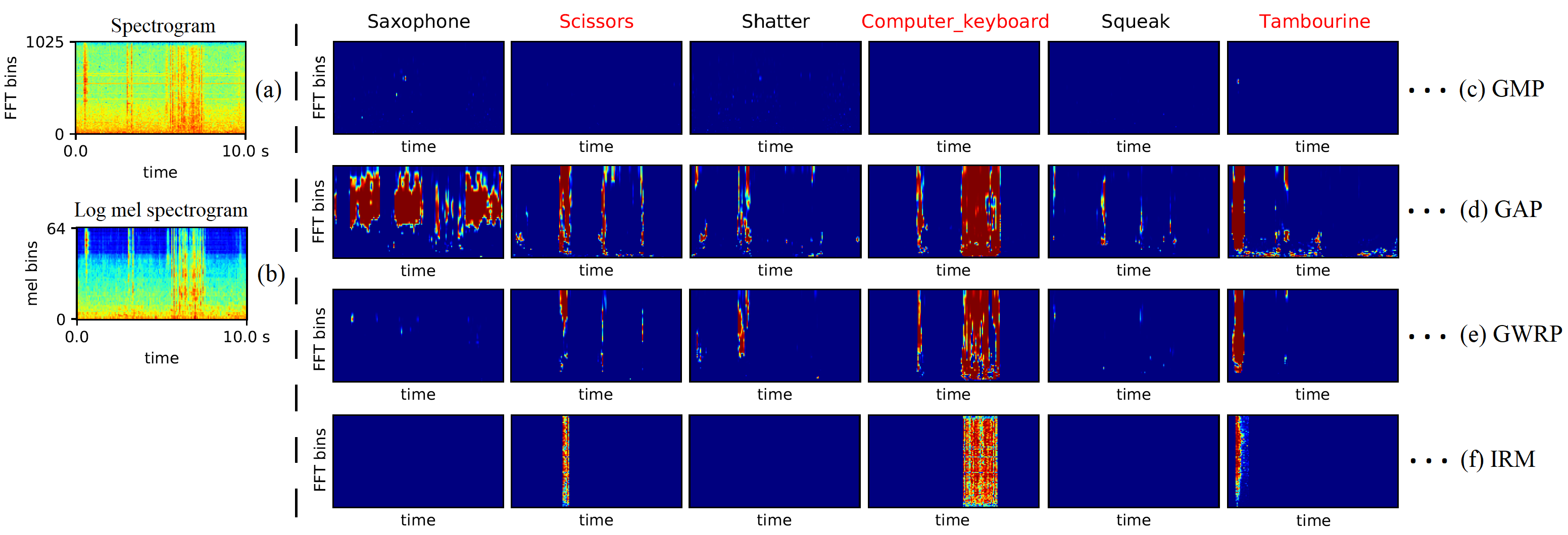}
	\caption{(a) Spectrogram of an audio clip containing ``scissors'', ``computer keyboard'' and ``tambourine'' (plotted in log scale); (b) Log mel spectrogram of the audio clip; (c) Upsampled T-F segmentation masks $ \widetilde{h}_{k} $ of sound events learned using global max pooling (GMP). Only a few T-F units have high value and the other parts of the T-F segmentation masks are dark; (d) Upsampled T-F segmentation masks $ \widetilde{h}_{k} $ of sound events learned using global average pooling (GAP); (e) Upsampled T-F segmentation masks $ \widetilde{h}_{k} $ of sound events learned using global weighted rank pooling (GWRP); (f) Ideal ratio mask (IRM) of sound events. Only 6 out of 41 T-F segmentation masks are plotted due to the limited space. }
	\label{fig:masks}
\end{figure*}
\subsection{Sound event separation}
As a byproduct, the T-F segmentation masks can be used to separate sound events from the mixture in the T-F domain. In addition, by applying an inverse Fourier transform on the separated T-F representation of each sound event, separated waveforms of the sound events can be obtained. Separating sound events from the mixture of sound events and background under a low SNR can improve the recognition of sound events in future work. Fig. \ref{fig:fig_ss} shows the pipeline of sound event separation. An audio clip $ x $ is presented to the segmentation mapping $ g_{1} $ to obtain T-F segmentation masks. Meanwhile, the complex spectrum $ \widetilde{X} $ of the audio clip is calculated. We use the tilde on $ X $ to distinguish the complex spectrum $ \widetilde{X} $ from the input T-F representation $ X $ because $ X $ might not be a spectrum, such as log mel spectrogram. We interpolate the segmentation masks of the input T-F representation $ h_{k}, k=1, ..., K $ to $ \widetilde{h}_{k}, k=1, ..., K $ representing the T-F segmentation masks of the complex spectrum. The reason for performing this interpolation is that $ \widetilde{h}_{k} $ may have a size different from $ h_{k} $, for example, a log mel spectrogram has fewer frequency bins than linear spectrum in the frequency domain. Then we multiply the upsampled T-F segmentation masks $ \widetilde{h}_{k} $ with the magnitude of the spectrum to obtain the segmented spectrogram of the $k$-th event:
\begin{equation} \label{eq:cmplx_x}
\widetilde{Y}_{k} = \widetilde{h}_{k} \odot \left | \widetilde{X} \right |, k=1, ..., K,
\end{equation}
\noindent where $ \odot $ represents the element-wise multiplication and $ \widetilde{Y}_{k} $ represents the segmented spectrogram of the $ k $-th event. Finally, an inverse Fourier transform with overlap add \cite{raki2005reducing} is applied on each segmented spectrogram with the phase from $ \widetilde{X} $ to obtain the separated waveforms $ \widehat{s}_{k}, k=1, ..., K $: 
\begin{equation} \label{eq:ifft}
\widehat{s}_{k} = \text{IFFT} \left ( \widetilde{Y}_{k} \cdot e^{j\angle \widetilde{X}} \right ). 
\end{equation}
We summarize the training, time-frequency segmentation, sound event detection and separation framework in Fig. \ref{fig:framework}. The training stage, sound event detection stage and sound event separation stage are shown in the left, middle and right column of Fig. \ref{fig:framework}, respectively. 

\section{Proposed segmentation mapping and classification mapping} \label{section:implementation}
In this section, we describe the implementation details of the segmentation mapping $ g_{1} $ and the classification mapping $ g_{2} $ proposed in Section \ref{section:wld}. 

\subsection{Segmentation mapping}\label{section:implementation_g1}
Segmentation mapping $ g_{1} $ takes a T-F representation of an audio clip as input and outputs segmentation masks of each sound event. We use log mel spectrogram as the input T-F representation, which has been shown to perform well in audio classification \cite{choi2016automatic, xu2017large, hershey2016cnn}. Ideally, the outputs of $ g_{1} $ are ideal ratio masks (IRMs) \cite{radfar2007single} of sound events in the T-F domain. The segmentation mapping $ g_{1} $ is modeled by a CNN. Each convolutional layer consists of a linear convolution, a batch normalization (BN) \cite{ioffe2015batch} and a ReLU \cite{nair2010rectified} nonlinearity as in \cite{ioffe2015batch}. The BN inserted between the convolution and the nonlinearity can stabilize and speed up the training \cite{ioffe2015batch}. We do not apply downsampling layers after convolutional layers because we want to retain the resolution of the input T-F segmentation masks. The T-F segmentation masks are obtained from the activations of the last CNN layer using a sigmoid non-linearity to constrain the values of the T-F segmentation masks to be between 0 and 1 to be a valid value of an IRM. The configuration details of the CNN will be described in Section \ref{section:experiment_model}.

The idea of learning the T-F segmentation masks explicitly is inspired by work on weakly labelled image localization \cite{zhou2016learning} and image segmentation \cite{simonyan2013deep, pathak2015constrained}. In weakly labelled image localization, saliency maps are learned indicating the locations of the objects in an image \cite{zhou2016learning}. Similarly, the T-F segmentation masks in our work resemble the saliency maps of an image \cite{zhou2016learning}, where T-F segmentation masks indicate what time and frequency a sound event occurs in a T-F representation.

\subsection{Classification mapping}\label{section:implementation_g2}
As described in Section \ref{section:wld}, the classification mapping $ g_{1} $ maps each segmentation mask $ h_{k} $ to the presence probability of its corresponding sound event. Modeling the classification mapping in different ways will lead to different representation of the segmentation masks (Fig. \ref{fig:masks}). We explored global max pooling \cite{choi2016automatic}, global average pooling \cite{lin2013network} and global rank pooling \cite{kolesnikov2016seed} for modeling the classification mappings $ g_{2} $. 

\subsubsection{Global max pooling}
Global max pooling (GMP) applied on feature maps has been used in audio tagging \cite{choi2016automatic}. GMP on each T-F segmentation mask map $ h_{k} $ is depicted as:
\begin{equation} \label{eq1_gmp}
F(h_{k}) = \underset{t,f}{\max} ~ h_{k}(t,f).
\end{equation}
GMP is based on the assumption that an audio clip contains a sound event if at least one T-F unit of the T-F input representation contains a sound event. GMP is invariant to the location of sound event in the T-F domain because whenever a sound event occurs, GMP will only select the maximum value of a T-F segmentation mask which is robust to the time or frequency shifts of the sound event. However, in the training stage, back propagation will only pass through the maximum value, so only a small part of data in the T-F domain are used to update the parameters in the neural network. Because of the maximum selection strategy, GMP encourages only one point in a T-F segmentation mask to be positive, so GMP will underestimate \cite{kolesnikov2016seed} the sound events in the T-F representation. Examples of T-F segmentation masks learned using GMP are shown in Fig. \ref{fig:masks}(c). 

\subsubsection{Global average pooling}
Global average pooling (GAP) was first applied in image classification \cite{lin2013network}. GAP on each T-F segmentation mask $ h_{k} $ is depicted as:
\begin{equation} \label{eq1_gap}
F(h_{k}) = \frac{1}{TF} \sum_{t}^{T} \sum_{f}^{F}h_{k}(t, f).
\end{equation}
GAP corresponds to the collective assumption in MIL \cite{amores2013multiple}, which states that all T-F units in a T-F segmentation mask contribute equally to the label of an audio clip. That is, all T-F units in a T-F segmentation mask are assumed to contain the labelled sound events. However, some sound events only last a short time, so GAP usually overestimates the sound events \cite{lin2013network}. Examples of T-F segmentation masks learned using GAP are shown in Fig \ref{fig:masks}(d). 

\subsubsection{Global weighted rank pooling}
To overcome the limitations of GMP and GAP, which underestimate and overestimate the sound events in the T-F segmentation masks, global weighted rank pooling (GWRP) is proposed in \cite{kolesnikov2016seed}. GWRP can be seen as a generalization of GMP and GAP. The idea of GWRP is to put a descending weight on the values of a T-F segmentation mask sorted in a descending order. Let an index set $ I^{c} = \{i_{1}, ... i_{M}\} $ define the descending order of the values within a T-F segmentation mask $ h_{k} $, i.e. $ (h_{k})_{i_{1}} \geq (h_{k})_{i_{2}} \geq ... \geq (h_{k})_{i_{n}} $, where $ M = T \times F $ is the number of T-F units in a T-F segmentation mask. Then the GWRP is defined as: 
\begin{equation} \label{eq1_gwrp}
F(h_{k}) = \frac{1}{Z(r)} \sum_{j=1}^{M} r^{j-1} (h_{k})_{i_{j}},
\end{equation}
\noindent where $ 0 \leq r \leq 1 $ is a hyper parameter and $ Z(r) = \sum_{j=1}^{M} r^{j-1} $ is a normalization term. When $ r=0 $ GWRP becomes GMP and when $ r=1 $ GWRP becomes GAP. The hyperparameter $ r $ can vary depending on the frequency of occurrence of the sound events. GWAP attends more to the T-F units of high values in a T-F segmentation mask and less to those of low values in a T-F segmentation mask. The T-F segmentation masks learned using GWMP is shown in Fig. \ref{fig:masks}(e). The ideal binary masks (IBMs) of the sound events are plotted in Fig. \ref{fig:masks}(f) for comparison with the GMP, GAP and GWRP. 

\subsection{Post-processing for sound event detection}\label{section:implementation_sed}
In Section \ref{section:framework} we mentioned that the frame-wise scores $ v_{k}(t) $ can be obtained from the T-F segmentation masks using Equation (\ref{eq:sed}). To reduce the number of false alarms, for an audio clip, we only apply sound event detection on the sound classes with positive audio tagging predictions. Then we apply thresholds on the frame-wise predictions $ v_{k}(t) $ to obtain the event-wise predictions. We apply a high threshold of 0.2 to detect the presence of sound events and then extend the boundary of both onset and offset sides until the frame-wise scores drop below threshold of 0.1. This two-step threshold method will produce smooth predictions of sound events. As the duration of sound events in DCASE 2018 Task 2 varies from 300 ms to 30 s, we remove the detected sound events that are shorter than 320 ms (10 frames) to reduce false alarms and join the sound events whose silence gap is shorter than 320 ms (10 frames).

\section{Experiments} \label{section:experiments}
\subsection{Dataset}
We mix the DCASE 2018 Task 1 acoustic scene dataset \cite{dcase2018task1} with the DCASE 2018 Task 2 general-purpose Freesound dataset \cite{dcase2018task2} under different signal-to-noise ratios (SNRs) to evaluate the proposed methods. The reason for this choice is that DCASE 2018 Task 1 provides background sounds recorded from a variety of real world scenes whereas the DCASE 2018 Task 2 provides a variety of foreground sound events. The DCASE 2018 Task 1 contains 8640 10-second audio clips in the development set of subtask A. The audio clips are recorded from 10 different scenes such as ``airport'', ``metro station'' and ``urban park''. The DCASE 2018 Task 2 contains 3710 manually verified sound events ranging in length from 300 ms to 30 s depending on the audio classes. There are 41 classes of sound events such as ``flute'', ``applause'' and ``cough''. We only use these manually verified audio clips from the DCASE 2018 Task 2 as sound events because the remaining audio clips are unverified and may contain noisy labels. We truncated the sound events to up to 2 seconds and mix them with the 10-second audio clips from the DCASE 2018 Task 1 acoustic scene dataset. The mixed audio clips are single channel with a sampling rate of 32 kHz. Each mixed audio clip contains three non-overlapped sound events. We mixed the sound events with the acoustic scenes for SNRs at 20dB, 10dB and 0dB. For each SNR, the 8000 mixed audio clips are divided into 4 cross-validation folds. Fig. \ref{fig:masks}(b) shows the log mel spectrogram of a mixed 10-second audio clip. The source code of our work is released\footnote{https://github.com/qiuqiangkong/sed\textunderscore time\textunderscore freq\textunderscore segmentation}.

\subsection{Evaluation metrics}
We use F-score \cite{mesaros2016metrics}, area under the curve (AUC) \cite{hanley1982meaning} and mean average precision (mAP) \cite{girshick2014rich} in the evaluation of the audio tagging, the frame-wise SED and the T-F segmentation. We also use error rate (ER) for evaluating the event-wise SED. 

\subsubsection{Basic statistics}
True positive (TP): Both the reference and the system prediction indicate an event to be active. False negative (FN): The reference indicates an event to be active but the system prediction indicates an event to be inactive. False positive (FP): The system prediction indicates an event to be active but the reference indicates it is not \cite{mesaros2016metrics}. 

\subsubsection{Precision, recall and F-score}
Precision (\textit{P}) and recall (\textit{R}) are defined as \cite{mesaros2016metrics}:
\begin{equation} \label{tp_fn_fp_tn}
P=\frac{TP}{TP+FP}, \qquad R=\frac{TP}{TP+FN}.
\end{equation}
Bigger \textit{P} and \textit{R} indicates better performance. F-score is calculated based on \textit{P} and \textit{R} \cite{mesaros2016metrics}:
\begin{equation} \label{prec_recall}
F=\frac{2P \cdot R}{P + R} = \frac{TP}{TP + (FN + FP)/2}.
\end{equation}
Bigger F-score indicates better performance. 

\subsubsection{Area under the curve (AUC)}
A receiver operating characteristic (ROC) curve \cite{hanley1982meaning} plots true positive rate (TPR) versus false positive rate (FPR). Area under the curve (AUC) score is the area under this ROC curve which summarizes the ROC curve to a single number. Using the AUC does not require manual selection of a threshold. Bigger AUC indicates better performance. A random guess has an AUC of 0.5. 

\subsubsection{Average precision}
Average precision (AP) is the average of the precision at different recall values. Similar to AUC, AP does not rely on the threshold. Different to AUC, AP does not count the true negatives and is widely used as a criterion in imbalanced dataset such as object detection \cite{girshick2014rich}.

\subsubsection{Error rate}
Error rate (ER) is an event-wise evaluation metric. ER measures the amount of errors in terms of \textit{insertions} (I), \textit{deletions} (D) and \textit{substitutions} (S) \cite{mesaros2016metrics}. For an audio clip, the insertions, deletions and substitutions are defined as: 
\begin{equation} \label{error_rate}
\begin{split}
S &=\text{min}(FN, FP), \\
D &=\text{max}(0, FN - FP), \\
I &=\text{max}(0, FP - FN),
\end{split}
\end{equation}
\noindent where FN, FP, FN are event-wise statistics in an audio clip. Lower \textit{ER}, \textit{S}, \textit{D} and \textit{I} indicate the better performance. When evaluating the event based criterion, we allow some degree of misalignment between a reference and a system output for counting a true positive \cite{mesaros2016metrics, mesaros2017dcase, mesaros2016tut}. Following the default configuration of \cite{mesaros2016metrics}, we adopt an onset collar of 200 ms and an offset collar of 200 ms / 50\% to count the true positive of a detection. We used the toolbox \cite{mesaros2016metrics} for evaluating the performance of the event-based SED.

\subsection{Feature extraction}
We apply a fast Fourier transform (FFT) with a window size of 2048 and an overlap of 1024 between neighbouring windows to extract the spectrogram of audio clips. This configuration that follows \cite{kong2018dcase} offers a good resolution in both time and frequency domain. Then mel filter banks with 64 bands are applied on the spectrogram followed by logarithm operation to obtain log mel spectrogram as the input T-F representation feature. Log mel spectrogram has been widely used in audio classification \cite{choi2016automatic, kong2018dcase}.
\begin{table}[]
\caption{Configuration of CNN. }
\label{table:configuration}
\begin{center}
{\renewcommand{\arraystretch}{1.5}
\begin{tabular}{c|c}
\hline
Layers & \thead{Output size \\ (feature maps $ \times $ time steps $\times$ mel bins)} \\ \hline
Input log mel spectrogram & $ 1 \times 311 \times 64 $  \\ \hline
$ \{ 3 \times 3, 32, \text{BN}, \text{ReLU} \} \times 2 $ & $ 32 \times 311 \times 64 $ \\ \hline
$ \{ 3 \times 3, 64, \text{BN}, \text{ReLU} \} \times 2 $ & $ 64 \times 311 \times 64 $  \\ \hline
$ \{ 3 \times 3, 128, \text{BN}, \text{ReLU} \} \times 2 $ & $ 128 \times 311 \times 64 $  \\ \hline
$ \{ 3 \times 3, 128, \text{BN}, \text{ReLU} \} \times 2 $ & $ 128 \times 311 \times 64 $ \\ \hline
$ 1 \times 1, 41, \text{sigmoid} $ & $ 41 \times 311 \times 64 $  \\ \hline
Global pooling (GP) & $ 41 $  \\ \hline
\end{tabular}}
\end{center}
\end{table}
\begin{table}
  \caption{F1-score, AUC and mAP of audio tagging at different SNRs. }
  \vspace{6pt}
  \label{table:at_brief}
  \resizebox{\columnwidth}{!}{%
  \centering
  \begin{tabular}{l p{.6cm}p{.6cm}p{.6cm}p{.6cm}p{.6cm}p{.6cm}p{.6cm}p{.6cm}p{.6cm}}
    \toprule
    & \multicolumn{3}{c}{\textbf{20 dB}} & \multicolumn{3}{c}{\textbf{10 dB}} & \multicolumn{3}{c}{\textbf{0 dB}} \\
	\cmidrule(lr){2-4} \cmidrule(lr){5-7} \cmidrule(lr){8-10}
    Algorithms & F1 & AUC & mAP & F1 & AUC & mAP & F1 & AUC & mAP \\
    \midrule
 DNN \cite{kong2016deep} & 0.439 & 0.885 & 0.468 & 0.396 & 0.861 & 0.402 & 0.331 & 0.810 & 0.314 \\
 WLD CNN \cite{kumar2017deep} & 0.498 & 0.777 & 0.498 & 0.524 & 0.794 & 0.526 & 0.528 & 0.815 & 0.535 \\
 FrameCNN \cite{2017_Chou_FrameCNN} & 0.581 & 0.899 & 0.587 & 0.543 & 0.883 & 0.526 & 0.484 & 0.850 & 0.439  \\
 Attention \cite{xu2017large} & \textbf{0.714} & 0.922 & \textbf{0.755} & \textbf{0.690} & 0.907 & \textbf{0.729} & \textbf{0.612} & 0.875 & \textbf{0.643} \\
 \midrule
 GMP & 0.435 & 0.818 & 0.475 & 0.406 & 0.801 & 0.440 & 0.373 & 0.773 & 0.389 \\
 GAP & 0.529 & 0.934 & 0.623 & 0.467 & 0.914 & 0.555 & 0.385 & 0.877 & 0.442 \\
 GWRP & 0.635 & \textbf{0.955} & 0.753 & 0.604 & \textbf{0.942} & 0.696 & 0.534 & \textbf{0.915} & 0.596 \\
	\bottomrule
\end{tabular}}
\end{table}
\begin{table*}[t]
\centering
\caption{F1-score of audio tagging at 0 dB SNR. }
\label{table:at_whole}
\resizebox{\textwidth}{!}{%
\begin{tabular}{l p{0.7cm}p{0.6cm}p{0.6cm}p{0.6cm}p{0.6cm}p{0.6cm}p{0.6cm}p{0.6cm}p{0.6cm}p{0.6cm}p{0.6cm}p{0.6cm}p{0.6cm}p{0.6cm}p{0.7cm}p{0.6cm}p{0.6cm}p{0.6cm}p{0.6cm}p{0.6cm}p{0.6cm}p{0.6cm}}
 \toprule
& Acous. guitar & Appla-use & Bark & Bass drum & Burp-ing & Bus & Cello & Chime & Clari-net & Keybo-ard & Cough & Cow-bell & Double bass & Drawer & Elec. piano & Fart & Finger snap &  Fire-works & Flute & Glock-enspiel & Gong\\
 \midrule
 DNN \cite{kong2016deep} & 0.286 & 0.873 & 0.332 & 0.041 & 0.344 & 0.367 & 0.489 & 0.546 & 0.423 & 0.283 & 0.075 & 0.133 & 0.197 & 0.083 & 0.304 & 0.267 & 0.389 & 0.285 & 0.350 & 0.464 & 0.310 \\
 WLD CNN \cite{kumar2017deep} & 0.633 & 0.896 & 0.719 & 0.547 & 0.794 & 0.248 & 0.610 & 0.589 & 0.504 & 0.390 & 0.513 & 0.889 & 0.436 & 0.136 & 0.435 & 0.384 & 0.672 & 0.375 & 0.270 & 0.692 & 0.513 \\
 FrameCNN \cite{2017_Chou_FrameCNN} & 0.416 & 0.878 & 0.719 & 0.166 & 0.557 & 0.385 & 0.529 & 0.562 & 0.448 & 0.507 & 0.484 & 0.668 & 0.314 & 0.181 & 0.392 & 0.304 & 0.556 & 0.474 & 0.385 & 0.488 & 0.465 \\
 Attention \cite{xu2017large} & 0.548 & 0.893 & 0.761 & 0.632 & 0.866 & 0.335 & 0.616 & 0.607 & 0.568 & 0.497 & 0.565 & 0.924 & 0.477 & 0.160  & 0.546 & 0.598 & 0.823 & 0.463 & 0.565 & 0.901 & 0.617 \\
 \midrule
 GMP & 0.458 & 0.522 & 0.335 & 0.183 & 0.400 & 0.087 & 0.299 & 0.468 & 0.424 & 0.422 & 0.151 & 0.774 & 0.281 & 0.076 & 0.279 & 0.284 & 0.176 & 0.271 & 0.315 & 0.844 & 0.434 \\
 GAP & 0.547 & 0.817 & 0.409 & 0.070 & 0.484 & 0.205 & 0.435 & 0.501 & 0.354 & 0.504 & 0.347 & 0.314 & 0.181 & 0.164 & 0.218 & 0.407 & 0.399 & 0.346 & 0.343 & 0.496 & 0.305 \\
 GWRP & 0.552 & 0.825 & 0.654 & 0.204 & 0.578 & 0.342 & 0.416 & 0.628 & 0.424 & 0.573 & 0.543 & 0.579 & 0.333 & 0.320 & 0.421 & 0.618 & 0.473 & 0.558 & 0.427 & 0.726 & 0.550 \\
 \bottomrule 
 \toprule 
 & Gunshot & Harmo-nica & Hi-hat & Keys & Knock & Laugh-ter & Meow & Micro-wave & Oboe & Saxo-phone & Sciss-ors & Shatter & Snare drum & Squeak & Tambo-urine & Tear-ing & Tele-phone & Trumpet & Violin & Writ-ing & Avg. \\
 \midrule
 DNN \cite{kong2016deep} & 0.297 & 0.672 & 0.547 & 0.418 & 0.276 & 0.192 & 0.075 & 0.121 & 0.408 & 0.500 & 0.411 & 0.336 & 0.368 & 0.097 & 0.299 & 0.254 & 0.270 & 0.528 & 0.379 & 0.293 & 0.331 \\
 WLD CNN \cite{kumar2017deep} & 0.538 & 0.742 & 0.910 & 0.643 & 0.649 & 0.361 & 0.359 & 0.263 & 0.589 & 0.636 & 0.558 & 0.410 & 0.599 & 0.052 & 0.593 & 0.436 & 0.324 & 0.642 & 0.755 & 0.349 & 0.528 \\
 FrameCNN \cite{2017_Chou_FrameCNN} & 0.424 & 0.723 & 0.688 & 0.660 & 0.553 & 0.390 & 0.355 & 0.400 & 0.490 & 0.528 & 0.497 & 0.481 & 0.624 & 0.193 & 0.733 & 0.449 & 0.346 & 0.526 & 0.475 & 0.431 & 0.484 \\
 Attention \cite{xu2017large} & 0.607 & 0.759 & 0.938 & 0.744 & 0.738 & 0.444 & 0.499 & 0.441 & 0.560 & 0.678 & 0.660 & 0.693 & 0.709 & 0.113 & 0.957 & 0.593 & 0.434 & 0.368 & 0.784 & 0.400 & \textbf{0.612} \\
 \midrule
 GMP & 0.398 & 0.322 & 0.796 & 0.141 & 0.483 & 0.311 & 0.275 & 0.207 & 0.442 & 0.474 & 0.173 & 0.251 & 0.465 & 0.031 & 0.891 & 0.504 & 0.329 & 0.585 & 0.567 & 0.175 & 0.373 \\
 GAP & 0.438 & 0.681 & 0.641 & 0.392 & 0.402 & 0.480 & 0.203 & 0.172 & 0.372 & 0.408 & 0.404 & 0.392 & 0.335 & 0.161 & 0.412 & 0.348 & 0.341 & 0.579 & 0.349 & 0.408 & 0.385 \\
 GWRP & 0.523 & 0.714 & 0.798 & 0.606 & 0.524 & 0.563 & 0.547 & 0.353 & 0.487 & 0.534 & 0.452 & 0.653 & 0.585 & 0.260 & 0.857 & 0.583 & 0.508 & 0.639 & 0.516 & 0.452 & 0.534 \\
  \bottomrule 
\end{tabular}}
\end{table*}
\begin{table*}[t]
\centering
\caption{F1-score of frame-wise SED at 0 dB SNR. }
\label{table:sed_framewise_whole}
\resizebox{\textwidth}{!}{%
\begin{tabular}{l p{0.7cm}p{0.6cm}p{0.6cm}p{0.6cm}p{0.6cm}p{0.6cm}p{0.6cm}p{0.6cm}p{0.6cm}p{0.6cm}p{0.6cm}p{0.6cm}p{0.6cm}p{0.6cm}p{0.7cm}p{0.6cm}p{0.6cm}p{0.6cm}p{0.6cm}p{0.6cm}p{0.6cm}p{0.6cm}}
 \toprule
& Acous. guitar & Appla-use & Bark & Bass drum & Burp-ing & Bus & Cello & Chime & Clari-net & Keybo-ard & Cough & Cow-bell & Double bass & Drawer & Elec. piano & Fart & Finger snap &  Fire-works & Flute & Glock-enspiel & Gong\\
 \midrule
 DNN \cite{kong2016deep} & 0.191 & 0.746 & 0.239 & 0.009 & 0.317 & 0.306 & 0.373 & 0.495 & 0.295 & 0.202 & 0.036 & 0.050 & 0.123 & 0.038 & 0.233 & 0.207 & 0.156 & 0.195 & 0.214 & 0.291 & 0.212 \\
 WLD CNN \cite{kumar2017deep} & 0.113 & 0.466 & 0.159 & 0.052 & 0.292 & 0.044 & 0.318 & 0.298 & 0.223 & 0.100 & 0.142 & 0.111 & 0.097 & 0.020 & 0.078 & 0.078 & 0.085 & 0.085 & 0.042 & 0.095 & 0.037 \\
 FrameCNN \cite{2017_Chou_FrameCNN} & 0.294 & 0.741 & 0.585 & 0.07  & 0.411 & 0.299 & 0.441 & 0.480 & 0.342 & 0.421 & 0.370 & 0.283 & 0.178 & 0.102 & 0.310 & 0.239 & 0.236 & 0.325 & 0.246 & 0.315 & 0.308 \\
 Attention \cite{xu2017large} & 0.062 & 0.422 & 0.069 & 0.020 & 0.189 & 0.024 & 0.242 & 0.263 & 0.210 & 0.019 & 0.059 & 0.051 & 0.045 & 0.003 & 0.068 & 0.050 & 0.076 & 0.031 & 0.159 & 0.026 & 0.088 \\
 \midrule
 GMP & 0.000 & 0.000 & 0.000 & 0.000 & 0.000 & 0.000 & 0.000 & 0.000 & 0.000 & 0.000 & 0.000 & 0.000 & 0.000 & 0.000 & 0.000 & 0.000 & 0.000 & 0.000 & 0.000 & 0.000 & 0.000 \\
 GAP & 0.410 & 0.661 & 0.338 & 0.033 & 0.341 & 0.139 & 0.240 & 0.429 & 0.195 & 0.426 & 0.269 & 0.121 & 0.088 & 0.108 & 0.170 & 0.297 & 0.102 & 0.229 & 0.173 & 0.214 & 0.200 \\
 GWRP & 0.453 & 0.704 & 0.507 & 0.072 & 0.456 & 0.188 & 0.326 & 0.575 & 0.341 & 0.457 & 0.402 & 0.222 & 0.193 & 0.172 & 0.351 & 0.498 & 0.247 & 0.355 & 0.316 & 0.596 & 0.418 \\
 \bottomrule 
 \toprule 
 & Gunshot & Harmo-nica & Hi-hat & Keys & Knock & Laugh-ter & Meow & Micro-wave & Oboe & Saxo-phone & Sciss-ors & Shatter & Snare drum & Squeak & Tambo-urine & Tear-ing & Tele-phone & Trumpet & Violin & Writ-ing & Avg. \\
 \midrule
 DNN \cite{kong2016deep} & 0.155 & 0.594 & 0.510 & 0.367 & 0.16  & 0.111 & 0.022 & 0.095 & 0.314 & 0.317 & 0.277 & 0.254 & 0.290 & 0.045 & 0.166 & 0.144 & 0.190 & 0.411 & 0.166 & 0.212 & 0.237 \\
 WLD CNN \cite{kumar2017deep} & 0.093 & 0.333 & 0.135 & 0.160 & 0.149 & 0.086 & 0.056 & 0.058 & 0.132 & 0.234 & 0.150 & 0.075 & 0.141 & 0.003 & 0.195 & 0.055 & 0.123 & 0.287 & 0.258 & 0.067 & 0.140 \\
 FrameCNN \cite{2017_Chou_FrameCNN} & 0.259 & 0.595 & 0.639 & 0.495 & 0.354 & 0.271 & 0.228 & 0.284 & 0.399 & 0.329 & 0.379 & 0.364 & 0.453 & 0.111 & 0.443 & 0.277 & 0.237 & 0.407 & 0.228 & 0.299 & 0.343 \\
 Attention \cite{xu2017large} & 0.029 & 0.143 & 0.107 & 0.096 & 0.101 & 0.051 & 0.034 & 0.018 & 0.137 & 0.353 & 0.078 & 0.038 & 0.054 & 0.005 & 0.188 & 0.046 & 0.148 & 0.08  & 0.156 & 0.056 & 0.100 \\
 \midrule
 GMP & 0.000 & 0.000 & 0.000 & 0.000 & 0.000 & 0.000 & 0.000 & 0.000 & 0.000 & 0.000 & 0.000 & 0.000 & 0.000 & 0.000 & 0.000 & 0.000 & 0.000 & 0.000 & 0.000 & 0.000 & 0.000 \\
 GAP & 0.231 & 0.528 & 0.553 & 0.332 & 0.233 & 0.294 & 0.133 & 0.121 & 0.237 & 0.167 & 0.146 & 0.319 & 0.265 & 0.114 & 0.191 & 0.274 & 0.172 & 0.437 & 0.105 & 0.313 & 0.252\\
 GWRP & 0.362 & 0.649 & 0.696 & 0.539 & 0.354 & 0.429 & 0.400 & 0.182 & 0.404 & 0.440 & 0.384 & 0.471 & 0.373 & 0.173 & 0.591 & 0.378 & 0.420 & 0.528 & 0.331 & 0.360 & \textbf{0.398} \\
  \bottomrule 
\end{tabular}}
\end{table*}
\begin{table}
  \caption{F1-score, AUC and mAP of frame-wise SED at different SNRs. }
  \vspace{6pt}
  \label{table:sed_framewise_brief}
  \resizebox{\columnwidth}{!}{%
  \centering
  \begin{tabular}{l p{.6cm}p{.6cm}p{.6cm}p{.6cm}p{.6cm}p{.6cm}p{.6cm}p{.6cm}p{.6cm}}
    \toprule
    & \multicolumn{3}{c}{\textbf{20 dB}} & \multicolumn{3}{c}{\textbf{10 dB}} & \multicolumn{3}{c}{\textbf{0 dB}} \\
	\cmidrule(lr){2-4} \cmidrule(lr){5-7} \cmidrule(lr){8-10}
    Algorithms & F1 & AUC & mAP & F1 & AUC & mAP & F1 & AUC & mAP \\
    \midrule
 DNN \cite{kong2016deep} & 0.360 & 0.722 & 0.269 & 0.306 & 0.702 & 0.224 & 0.237 & 0.666 & 0.169 \\
 WLD CNN \cite{kumar2017deep} & 0.168 & 0.669 & 0.179 & 0.182 & 0.688 & 0.201 & 0.140 & 0.701 & 0.166 \\
 FrameCNN \cite{2017_Chou_FrameCNN} & 0.440 & 0.808 & 0.369 & 0.399 & 0.787 & 0.329 & 0.343 & 0.756 & 0.275 \\
 Attention \cite{xu2017large} & 0.163 & 0.827 & 0.317 & 0.137 & 0.807 & 0.278 & 0.100 & 0.773 & 0.221 \\
 \midrule
 GMP & 0.000 & 0.676 & 0.090 & 0.000 & 0.658 & 0.076 & 0.000 & 0.649 & 0.072 \\
 GAP & 0.398 & 0.790 & 0.400 & 0.334 & 0.753 & 0.328 & 0.252 & 0.712 & 0.245 \\
 GWRP & \textbf{0.511} & \textbf{0.886} & \textbf{0.508} & \textbf{0.472} & \textbf{0.871} & \textbf{0.453} & \textbf{0.398} & \textbf{0.829} & \textbf{0.360} \\
	\bottomrule
\end{tabular}}
\end{table}
\begin{table}
  \caption{F1-score, AUC and mAP of event-wise SED at different SNRs. }
  \vspace{6pt}
  \label{table:sed_eventwise_brief}
  \resizebox{\columnwidth}{!}{%
  \centering
  \begin{tabular}{l p{0.45cm}p{0.45cm}p{0.45cm}p{0.45cm}p{0.45cm}p{0.45cm}p{0.45cm}p{0.45cm}p{0.45cm}p{0.45cm}p{0.45cm}p{0.45cm}}
    \toprule
    & \multicolumn{4}{c}{\textbf{20 dB}} & \multicolumn{4}{c}{\textbf{10 dB}} & \multicolumn{4}{c}{\textbf{0 dB}} \\
	\cmidrule(lr){2-5} \cmidrule(lr){6-9} \cmidrule(lr){10-13}
    Algorithms & F1 & ER & D & I & F1 & ER & D & I & F1 & ER & D & I \\
    \midrule
 DNN \cite{kong2016deep} & 0.226 & 1.91 & 0.75 & 1.16 & 0.178 & 2.29 & 0.79 & 1.50 & 0.120 & 2.80 & 0.84 & 1.96 \\
 WLD CNN \cite{kumar2017deep} & 0.010 & 1.16 & 0.99 & 0.17 & 0.011 & 1.15 & 0.99 & 0.17 & 0.018 & 1.12 & 0.99 & 0.13 \\
 FrameCNN \cite{2017_Chou_FrameCNN} & 0.166 & 2.38 & 0.79 & 1.58 & 0.151 & 2.49 & 0.81 & 1.68 & 0.141 & 2.70 & 0.81 & 1.88 \\
 Attention \cite{xu2017large} & 0.028 & 1.10 & 0.96 & 0.14 & 0.021 & 1.10 & 0.97 & 0.13 & 0.011 & 1.09 & 0.98 & 0.10 \\
 \midrule
 GMP & 0.000 & 1.00 & 1.00 & 0.00 & 0.000 & 1.00 & 1.00 & 0.00 & 0.000 & 1.00 & 1.00 & 0.00 \\
 GAP & 0.173 & 2.71 & 0.78 & 1.93 & 0.139 & 2.95 & 0.82 & 2.13 & 0.098 & 3.52 & 0.86 & 2.66 \\
 GWRP & \textbf{0.254} & 2.12 & 0.66 & 1.45 & \textbf{0.227} & 2.30 & 0.69 & 1.61 & \textbf{0.167} & 2.55 & 0.76 & 1.78 \\
	\bottomrule
\end{tabular}}
\end{table}
\begin{table*}[t]
\centering
\caption{F1-score of event-wise SED at 0 dB SNR. }
\label{table:sed_eventwise_whole}
\resizebox{\textwidth}{!}{%
\begin{tabular}{l p{0.7cm}p{0.6cm}p{0.6cm}p{0.6cm}p{0.6cm}p{0.6cm}p{0.6cm}p{0.6cm}p{0.6cm}p{0.6cm}p{0.6cm}p{0.6cm}p{0.6cm}p{0.6cm}p{0.7cm}p{0.6cm}p{0.6cm}p{0.6cm}p{0.6cm}p{0.6cm}p{0.6cm}p{0.6cm}}
 \toprule
& Acous. guitar & Appla-use & Bark & Bass drum & Burp-ing & Bus & Cello & Chime & Clari-net & Keybo-ard & Cough & Cow-bell & Double bass & Drawer & Elec. piano & Fart & Finger snap &  Fire-works & Flute & Glock-enspiel & Gong\\
 \midrule
 DNN \cite{kong2016deep} & 0.132 & 0.287 & 0.083 & 0.002 & 0.176 & 0.233 & 0.125 & 0.389 & 0.041 & 0.141 & 0.033 & 0.007 & 0.068 & 0.036 & 0.141 & 0.113 & 0.035 & 0.113 & 0.036 & 0.079 & 0.159 \\
 WLD CNN \cite{kumar2017deep} & 0.020 & 0.013 & 0.001 & 0.001 & 0.110 & 0.001 & 0.036 & 0.067 & 0.025 & 0.005 & 0.001 & 0.001 & 0.003 & 0.001 & 0.001 & 0.001 & 0.001 & 0.001 & 0.001 & 0.001 & 0.001 \\
 FrameCNN \cite{2017_Chou_FrameCNN} & 0.098 & 0.510 & 0.187 & 0.012 & 0.090 & 0.180 & 0.287 & 0.186 & 0.157 & 0.194 & 0.144 & 0.005 & 0.04  & 0.091 & 0.168 & 0.163 & 0.042 & 0.133 & 0.081 & 0.265 & 0.098 \\
 Attention \cite{xu2017large} & 0.000 & 0.051 & 0.000 & 0.000 & 0.052 & 0.000 & 0.020 & 0.018 & 0.000 & 0.000 & 0.000 & 0.000 & 0.009 & 0.000 & 0.000 & 0.000 & 0.000 & 0.000 & 0.019 & 0.000 & 0.000 \\
 \midrule
 GMP & 0.000 & 0.000 & 0.000 & 0.000 & 0.000 & 0.000 & 0.000 & 0.000 & 0.000 & 0.000 & 0.000 & 0.000 & 0.000 & 0.000 & 0.000 & 0.000 & 0.000 & 0.000 & 0.000 & 0.000 & 0.000 \\
 GAP & 0.060 & 0.416 & 0.141 & 0.000 & 0.150 & 0.035 & 0.123 & 0.178 & 0.085 & 0.296 & 0.107 & 0.016 & 0.033 & 0.070 & 0.025 & 0.197 & 0.003 & 0.078 & 0.054 & 0.000 & 0.032 \\
 GWRP & 0.131 & 0.225 & 0.315 & 0.002 & 0.352 & 0.030 & 0.086 & 0.363 & 0.086 & 0.211 & 0.228 & 0.010 & 0.089 & 0.111 & 0.153 & 0.312 & 0.068 & 0.144 & 0.060 & 0.004 & 0.281 \\
 \bottomrule 
 \toprule 
 & Gunshot & Harmo-nica & Hi-hat & Keys & Knock & Laugh-ter & Meow & Micro-wave & Oboe & Saxo-phone & Sciss-ors & Shatter & Snare drum & Squeak & Tambo-urine & Tear-ing & Tele-phone & Trumpet & Violin & Writ-ing & Avg. \\
 \midrule
 DNN \cite{kong2016deep} & 0.073 & 0.455 & 0.205 & 0.262 & 0.095 & 0.054 & 0.024 & 0.047 & 0.135 & 0.107 & 0.128 & 0.174 & 0.106 & 0.020 & 0.057 & 0.088 & 0.100 & 0.140 & 0.031 & 0.173 & 0.120 \\
 WLD CNN \cite{kumar2017deep} & 0.001 & 0.153 & 0.001 & 0.003 & 0.008 & 0.003 & 0.001 & 0.001 & 0.007 & 0.043 & 0.005 & 0.001 & 0.011 & 0.001 & 0.001 & 0.001 & 0.073 & 0.077 & 0.063 & 0.001& 0.018 \\
 FrameCNN \cite{2017_Chou_FrameCNN} & 0.044 & 0.226 & 0.409 & 0.142 & 0.071 & 0.113 & 0.140 & 0.120 & 0.223 & 0.077 & 0.140 & 0.134 & 0.132 & 0.042 & 0.031 & 0.104 & 0.071 & 0.241 & 0.052 & 0.124 & 0.141 \\
 Attention \cite{xu2017large} & 0.000 & 0.000 & 0.000 & 0.000 & 0.106 & 0.000 & 0.000 & 0.000 & 0.000 & 0.188 & 0.000 & 0.000 & 0.000 & 0.000 & 0.000 & 0.000 & 0.000 & 0.000 & 0.000 & 0.000 & 0.011 \\
 \midrule
 GMP & 0.000 & 0.000 & 0.000 & 0.000 & 0.000 & 0.000 & 0.000 & 0.000 & 0.000 & 0.000 & 0.000 & 0.000 & 0.000 & 0.000 & 0.000 & 0.000 & 0.000 & 0.000 & 0.000 & 0.000 & 0.000 \\
 GAP & 0.035 & 0.440 & 0.002 & 0.103 & 0.095 & 0.152 & 0.062 & 0.064 & 0.153 & 0.024 & 0.041 & 0.024 & 0.075 & 0.078 & 0.003 & 0.078 & 0.038 & 0.257 & 0.013 & 0.184 & 0.098 \\
 GWRP & 0.078 & 0.519 & 0.031 & 0.356 & 0.206 & 0.205 & 0.269 & 0.118 & 0.252 & 0.165 & 0.167 & 0.130 & 0.065 & 0.067 & 0.065 & 0.146 & 0.238 & 0.175 & 0.105 & 0.243 & \textbf{0.167} \\
  \bottomrule 
\end{tabular}}
\end{table*}
\begin{table*}[t]
\centering
\caption{F1-score of time-frequency segmentation at 0 dB SNR. }
\label{table:ss_whole}
\resizebox{\textwidth}{!}{%
\begin{tabular}{l p{0.7cm}p{0.6cm}p{0.6cm}p{0.6cm}p{0.6cm}p{0.6cm}p{0.6cm}p{0.6cm}p{0.6cm}p{0.6cm}p{0.6cm}p{0.6cm}p{0.6cm}p{0.6cm}p{0.7cm}p{0.6cm}p{0.6cm}p{0.6cm}p{0.6cm}p{0.6cm}p{0.6cm}p{0.6cm}}
 \toprule
& Acous. guitar & Appla-use & Bark & Bass drum & Burp-ing & Bus & Cello & Chime & Clari-net & Keybo-ard & Cough & Cow-bell & Double bass & Drawer & Elec. piano & Fart & Finger snap &  Fire-works & Flute & Glock-enspiel & Gong\\
 \midrule
 GMP & 0.000 & 0.001 & 0.001 & 0.000 & 0.002 & 0.000 & 0.003 & 0.002 & 0.002 & 0.002 & 0.000 & 0.005 & 0.001 & 0.000 & 0.001 & 0.000 & 0.000 & 0.001 & 0.001 & 0.002 & 0.002 \\
 GAP & 0.128 & 0.391 & 0.106 & 0.009 & 0.155 & 0.073 & 0.124 & 0.187 & 0.057 & 0.201 & 0.143 & 0.038 & 0.044 & 0.068 & 0.067 & 0.126 & 0.029 & 0.119 & 0.052 & 0.081 & 0.116 \\
 GWRP & 0.222 & 0.519 & 0.226 & 0.030 & 0.291 & 0.095 & 0.213 & 0.313 & 0.114 & 0.303 & 0.241 & 0.125 & 0.086 & 0.100 & 0.127 & 0.256 & 0.092 & 0.204 & 0.104 & 0.212 & 0.237 \\
 \bottomrule 
 \toprule 
 & Gunshot & Harmo-nica & Hi-hat & Keys & Knock & Laugh-ter & Meow & Micro-wave & Oboe & Saxo-phone & Sciss-ors & Shatter & Snare drum & Squeak & Tambo-urine & Tear-ing & Tele-phone & Trumpet & Violin & Writ-ing & Avg. \\
 \midrule
 GMP & 0.001 & 0.002 & 0.001 & 0.002 & 0.001 & 0.001 & 0.000 & 0.000 & 0.001 & 0.001 & 0.000 & 0.000 & 0.002 & 0.000 & 0.001 & 0.001 & 0.001 & 0.002 & 0.003 & 0.001 & 0.001 \\
 GAP & 0.139 & 0.264 & 0.212 & 0.139 & 0.074 & 0.135 & 0.085 & 0.055 & 0.077 & 0.120 & 0.085 & 0.144 & 0.108 & 0.082 & 0.057 & 0.140 & 0.059 & 0.166 & 0.074 & 0.130 & 0.114\\
 GWRP & 0.283 & 0.379 & 0.497 & 0.311 & 0.190 & 0.249 & 0.185 & 0.085 & 0.140 & 0.257 & 0.213 & 0.272 & 0.196 & 0.108 & 0.327 & 0.237 & 0.138 & 0.313 & 0.222 & 0.215 & \textbf{0.218} \\
  \bottomrule 
\end{tabular}}
\end{table*}
\begin{table}
  \caption{F1-score, AUC and mAP of time-frequency segmentation at different SNRs. }
  \vspace{6pt}
  \label{table:ss_brief}
  \resizebox{\columnwidth}{!}{%
  \centering
  \begin{tabular}{l p{.6cm}p{.6cm}p{.6cm}p{.6cm}p{.6cm}p{.6cm}p{.6cm}p{.6cm}p{.6cm}}
    \toprule
    & \multicolumn{3}{c}{\textbf{20 dB}} & \multicolumn{3}{c}{\textbf{10 dB}} & \multicolumn{3}{c}{\textbf{0 dB}} \\
	\cmidrule(lr){2-4} \cmidrule(lr){5-7} \cmidrule(lr){8-10}
    Algorithms & F1 & AUC & mAP & F1 & AUC & mAP & F1 & AUC & mAP \\
    \midrule
 GMP & 0.001 & 0.347 & 0.008 & 0.001 & 0.345 & 0.007 & 0.001 & 0.362 & 0.005 \\
 GAP & 0.215 & \textbf{0.889} & 0.230 & 0.168 & \textbf{0.880} & 0.187 & 0.114 & \textbf{0.861} & 0.143 \\
 GWRP & \textbf{0.324} & 0.849 & \textbf{0.268} & \textbf{0.280} & 0.845 & \textbf{0.227} & \textbf{0.218} & 0.836 & \textbf{0.175} \\
	\bottomrule
\end{tabular}}
\end{table}
\begin{figure}[t]
	\centering
	\label{fig:framewise_sed}
	\includegraphics[width=\columnwidth]{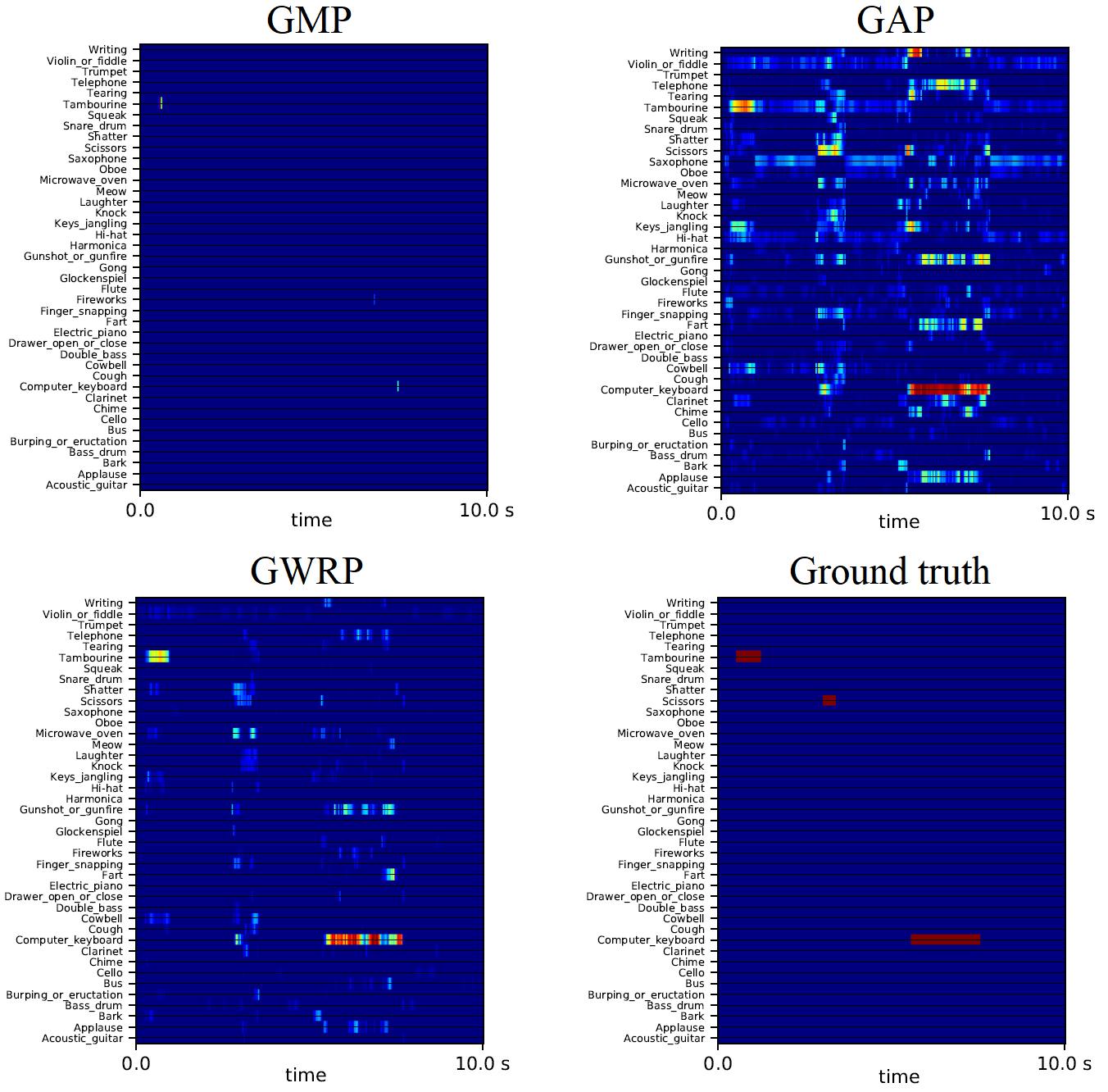}
	\caption{Frame-wise predictions using GMP, GAP, GWRP with SNR at 0 dB. The ground truth annotation is shown in the bottom right. }
	\label{fig:sed}
\end{figure}

\subsection{Model}\label{section:experiment_model}
In this subsection we give a detailed description of the configuration of the segmentation mapping in Section \ref{section:implementation_g1} and the classification mapping in Section \ref{section:implementation_g2}. We apply a ``VGG-like'' convolutional neural network \cite{simonyan2014very} with 8 convolutional blocks on the input log mel spectrogram \cite{kong2018dcase}. Each convolutional layer consists of a linear convolution with a filter size of $ 3 \times 3 $ followed by a batch normalization layer \cite{ioffe2015batch} and a ReLU activation function \cite{nair2010rectified}. We use 4 convolution blocks following the baseline system of DCASE 2018 \cite{kong2018dcase}. The number of feature maps of the convolutional layers are 32, 64, 128 and 128, respectively. This configuration is to fit the model to a single GPU card with 12 GB RAM sufficiently. Then a $ 1 \times 1 $ convolutional layer with sigmoid non-linearity is applied to convert the feature maps to the T-F segmentation masks of sound events. Then a global pooling is used to summarize each T-F segmentation mask to a scalar representing the presence probability of the sound events in an audio clip. We summarize the configuration of the neural network in Table \ref{table:configuration}. In training we use a mini-batch size of 24 to fully utilize the single card GPU with 12 GB RAM. The Adam optimizer \cite{kingma2014adam} with a learning rate 0.001 is used for its fast convergence.

\subsection{Audio tagging}
We compare our method with fully connected neural network \cite{kong2016deep}, CNN trained on weakly labelled data \cite{kumar2017deep}, FrameCNN \cite{2017_Chou_FrameCNN} and the attention model \cite{xu2017large}. We apply GMP, GAP and GWRP as global pooling in our model. Table \ref{table:at_brief} shows that for SNR at 20 dB, the attention model \cite{xu2017large} achieves the best F1-score of 0.714 and mAP of 0.755 followed by the GWRP of 0.635 and 0.753, respectively. On the other hand, GWRP achieves the best AUC of 0.955. Comparing the performance under different SNRs, the F1-score and mAP drop approximately 0.1 in absolute value for SNR changed from 20 dB to 0 dB. AUC drop approximately 0.04 in absolute value for SNR changed from 20 dB to 0 dB. This result shows that there is a large variance in audio tagging under low SNR. Table \ref{table:at_whole} shows the audio tagging results of all sound events under 0 dB SNR. Some sound events such as ``hi-hat'' and ``tambourine'' have higher classification accuracy while some sound events such as ``microwave'' and ``squeak'' are difficult to recognize. On average, the attention model \cite{xu2017large} achieves the best F1-score of 0.612 followed by GWRP of 0.534.

\subsection{Frame-wise sound event detection}
Table \ref{table:sed_framewise_whole} shows the F1-score of the frame-wise SED for all sound classes under SNR of 0 dB. GWRP achieves the best averaged F1-score of 0.398, followed by the FrameCNN model \cite{2017_Chou_FrameCNN} of 0.343. Some classes such as ``applause'' and ``hi-hat'' have higher F1-score by the frame-wise SED, while some classes such as ``drawer'' and squeak'' have lower F1-score by the frame-wise SED. Table \ref{table:sed_framewise_brief} shows the frame-wise SED results under different SNRs. GWRP achieves the best F1-score, AUC and mAP of 0.511, 0.886 and 0.508 under 20 dB SNR. The FrameCNN model \cite{2017_Chou_FrameCNN} achieves a second place with an F1-score of 0.440. GAP overestimates the sound events which is shown in the visualization of the upsampled T-F segmentation masks (Fig. 7). GAP does not perform better than GWRP. GMP underestimates the sound events (Fig. 7) and performs worst in frame-wise SED. In GWRP, the F1-score drops from 0.511 to 0.472 to 0.398 under SNRs of 20 dB, 10 dB and 0 dB. Fig. 8 shows the frame-wise scores of sound events obtained from equation (\ref{eq:sed}) under SNR of 0 dB. Frame-wise scores obtained by using GWRP looks closer to the ground truth than obtained using GMP and GAP. Compared with event-wise SED, frame-wise SED does not depend on post-processing. 

\subsection{Event-wise sound event detection}
Although frame-wise SED does not depend on post-processing so is a more objective criterion, it makes more sense to have event-wise predictions. The event-wise predictions are obtained from frame-wise predictions following Section \ref{section:implementation_sed}. Table \ref{table:sed_eventwise_brief} shows that the GWRP achieves the best F1-score of 0.254 in event-wise SED. Although GMP seems to achieve the lowest ER of 1.00, GMP deletes all the events and has a deletion error of 1.00 and an insertion of 0. On the other hand, GWRP has the lowest deletion error of 0.66 and has an insertion error of 1.45. The F1-scores drop from 0.254 to 0.227 to 0.167 under SNRs of 20 dB, 10 dB and 0 dB. Table \ref{table:sed_eventwise_whole} shows the the F1-score of event-wise SED of all sound classes. Some sound classes such as ``barks'', ``harmonica'' have higher detection F1-score. GWRP achieves the best averaged F1-score of 0.167.

\subsection{Time-frequency segmentation}
Table \ref{table:ss_whole} shows the T-F segmentation results of all sound classes under 0 dB. As the T-F segmentation can not be obtained by previous works including the fully connected neural network \cite{kong2016deep}, the CNN trained on weakly labelled data \cite{kumar2017deep}, the FrameCNN \cite{2017_Chou_FrameCNN} and the attention model \cite{xu2017large}, we only report the T-F segmentation results with our proposed methods. GWRP achieves the best F1-score of 0.218 on average. Table \ref{table:ss_brief} shows the T-F segmentation results under different SNRs. Table \ref{table:ss_brief} shows that GWRP achieves the best F1-score, AUC and mAP of 0.324, 0.849 and 0.268 under 20 dB SNR, respectively. GMP underestimates the T-F segmentation masks and performs the worst in T-F segmentation. GAP overestimates the T-F segmentation masks and performs worse than GWRP in F1-score. The T-F segmentation masks learned by GWRP (Fig. \ref{fig:masks}(e)) looks closer to the IRM than the T-F segmentation masks learned by using GMP and GAP.

\section{Conclusion} \label{section:conclusion}
This paper proposes a time-frequency (T-F) segmentation, sound event detection and separation framework trained on weakly labelled data. In training, a segmentation mapping and a classification mapping are trained jointly using the weakly labelled data. In T-F segmentation, we use the trained segmentation mapping to calculate the T-F segmentation masks. Detected sound events can then be obtained from the T-F segmentation masks. As a byproduct, separated waveforms of sound events can be obtained from the T-F segmentation masks. Experiments show that the global weighted rank pooling (GWRP) outperforms the global max pooling, the global average pooling and previously proposed systems in both of T-F segmentation and sound event detection. The limitation of this approach is that the T-F segmentation masks are not perfectly matching the ideal ratio mask (IRM) of the sound events. In future, we will improve the T-F segmentation masks to match the IRM for event separation.  

\section*{Acknowledgment}
This research was supported by EPSRC grant EP/N014111/1 ``Making Sense of Sounds'' and a Research Scholarship from the China Scholarship Council (CSC) No. 201406150082. Iwona Sobieraj is sponsored by the European Union's H2020 Framework Programme (H2020-MSCA-ITN-2014) under grant agreement No. 642685 MacSeNet. The authors thank Dominic Ward for helping to improve the paper in the early stage. The authors thank all anonymous reviewers for their effort and suggestions to improve this paper.

\ifCLASSOPTIONcaptionsoff
  \newpage
\fi



%

%

\bibliographystyle{unsrt}
\bibliographystyle{IEEEtran}
\bibliography{ref}

\begin{IEEEbiography}[{\includegraphics[width=1in,height=1.25in,clip,keepaspectratio]{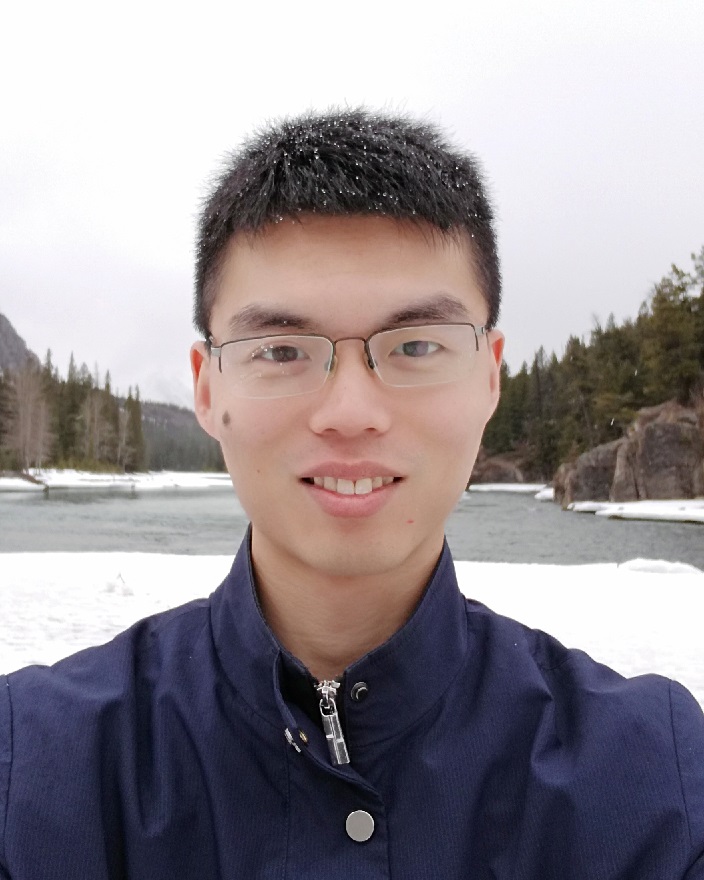}}]{Qiuqiang Kong}
(S'17) received the B.Sc. and the M.E. degree in South China University of Techology, Guangzhou, China, in 2012 and 2015, respectively. He is currently pursuing a PhD degree in University of Surrey, Guildford, UK. His research interest includes audio signal processing and machine learning. 
\end{IEEEbiography}

\begin{IEEEbiography}[{\includegraphics[width=1in,height=1.25in,clip,keepaspectratio]{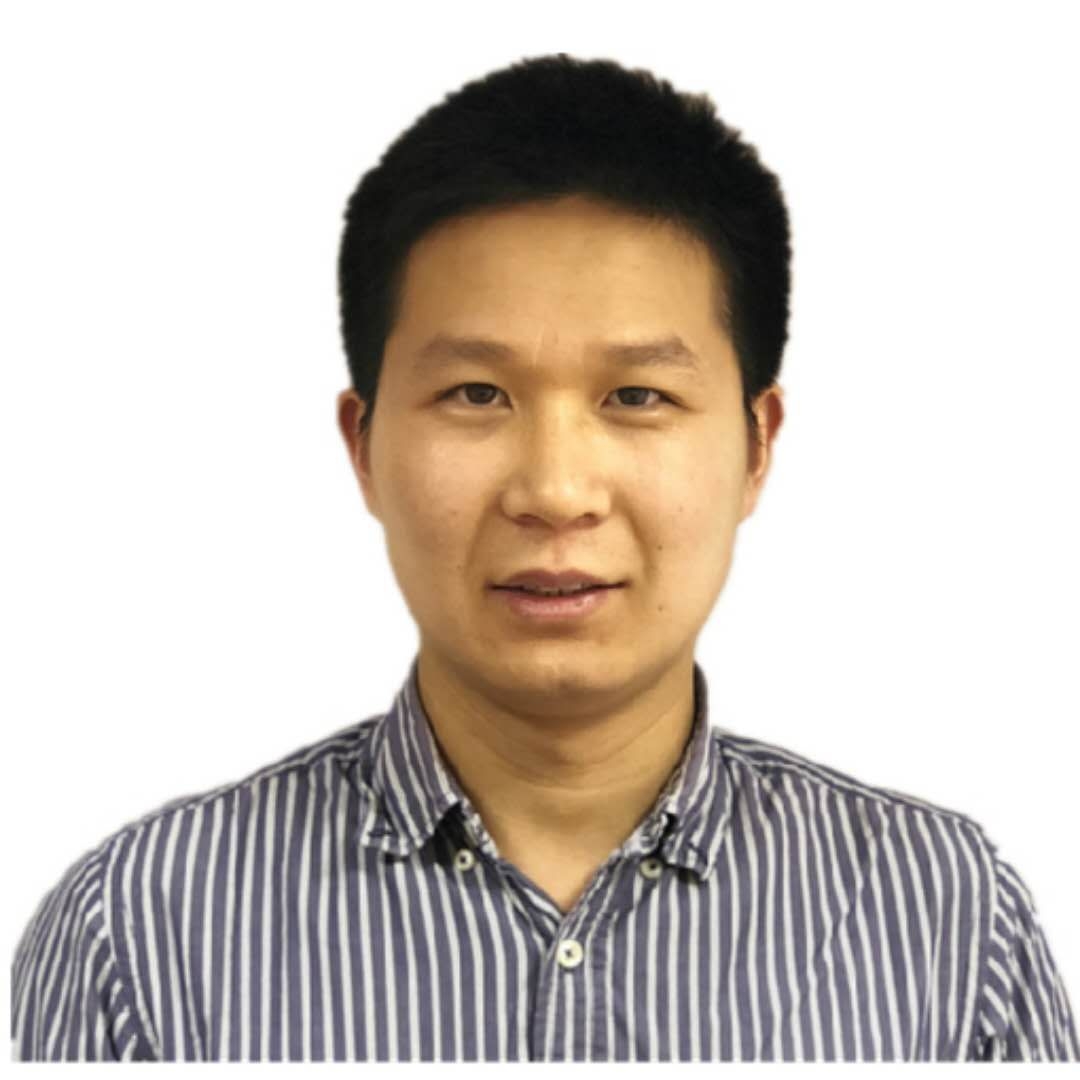}}]{Yong Xu}
(M'17) received the Ph.D. degree from the University of Science and Technology of China (USTC), Hefei, China, in 2015, on the topic of DNN-based speech enhancement and recognition. Currently, he is a senior research scientist in Tencent AI lab, Bellevue, USA.  He once worked at the University of Surrey, U.K. as a Research Fellow from 2016 to 2018 working on sound event detection. He visited Prof. Chin-Hui Lee's lab in Georgia Institute of Technology, USA from Sept. 2014 to May 2015. He once also worked in IFLYTEK company from 2015 to 2016 to develop far-field ASR technologies. His research interests include deep learning, speech enhancement and recognition, sound event detection, etc. He received 2018 IEEE SPS best paper award.
\end{IEEEbiography}

\begin{IEEEbiography}[{\includegraphics[width=1in,height=1.25in,clip,keepaspectratio]{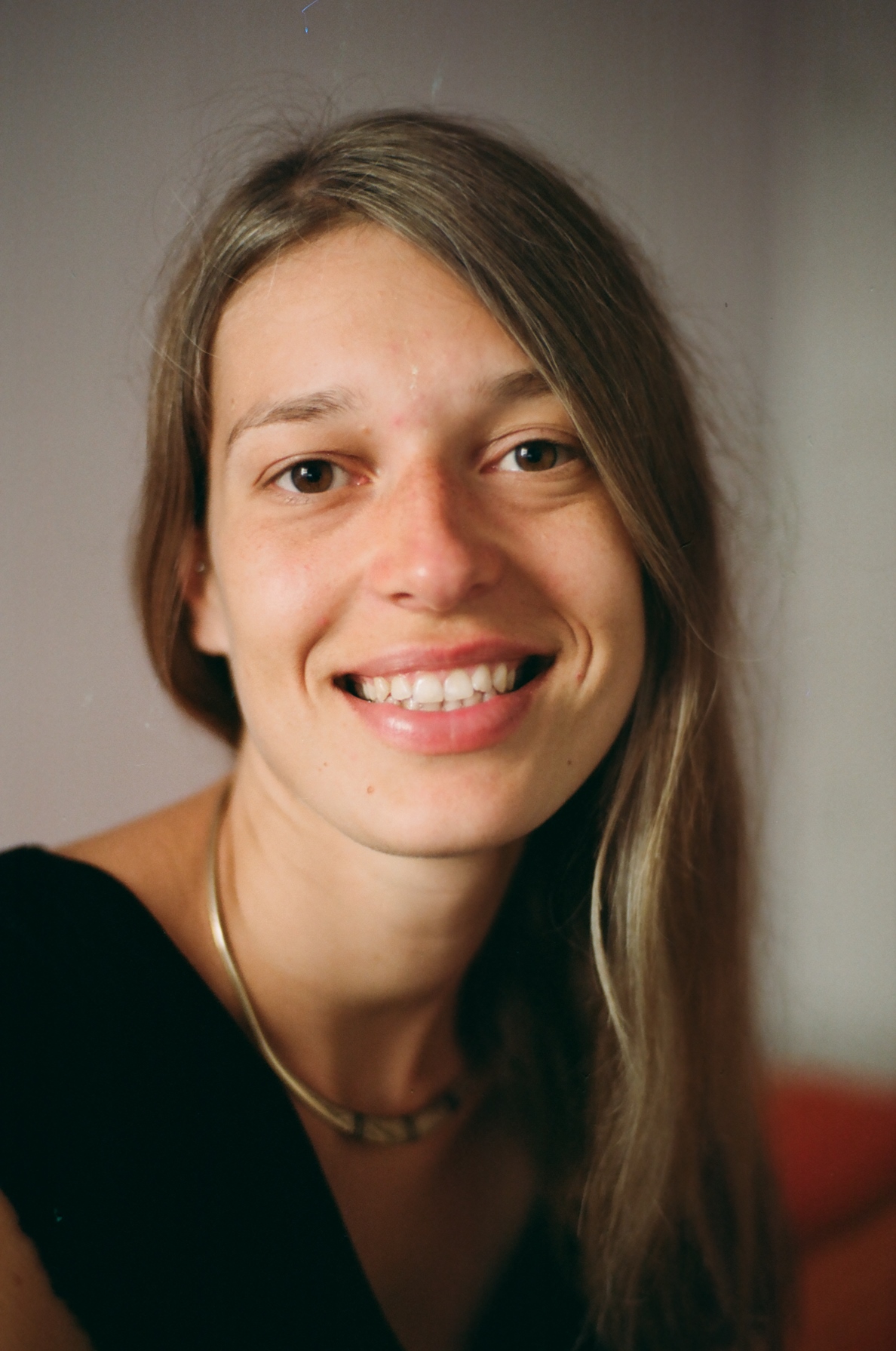}}]{Iwona Sobieraj}
 received the B.A. and the M.E. degreed from Warsaw University of Technology, Poland, in 2010 and 2011, respectively. She joined Samsung Electronics R\&D, Warsaw, Poland in 2012. Since 2015 she is pursuing a PhD degree at the University of Surrey, Guildford, UK. Her main research interest include environmental audio analysis, non-negative matrix factorization and deep learning.
\end{IEEEbiography}

\begin{IEEEbiography}[{\includegraphics[width=1in,height=1.25in,clip,keepaspectratio]{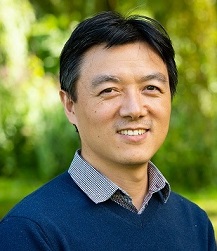}}]{Wenwu Wang}
(M'02-SM'11) was born in Anhui, China. He received the B.Sc. degree in 1997, the M.E. degree in 2000, and the Ph.D. degree in 2002, all from Harbin Engineering University, China. He then worked in King's College London, Cardiff University, Tao Group Ltd. (now Antix Labs Ltd.), and Creative Labs, before joining University of Surrey, UK, in May 2007, where he is currently a Reader in Signal Processing, and a Co-Director of the Machine Audition Lab within the Centre for Vision Speech and Signal Processing. He has been a Guest Professor at Qingdao University of Science and Technology, China, since 2018. His current research interests include blind signal processing, sparse signal processing, audio-visual signal processing, machine learning and perception, machine audition (listening), and statistical anomaly detection. He has (co)-authored over 200 publications in these areas. He served as an Associate Editor for IEEE Transactions on Signal Processing from 2014 to 2018. He is also Publication Co-Chair for ICASSP 2019, Brighton, UK. 

\end{IEEEbiography}

\begin{IEEEbiography}[{\includegraphics[width=1in,height=1.25in,clip,keepaspectratio]{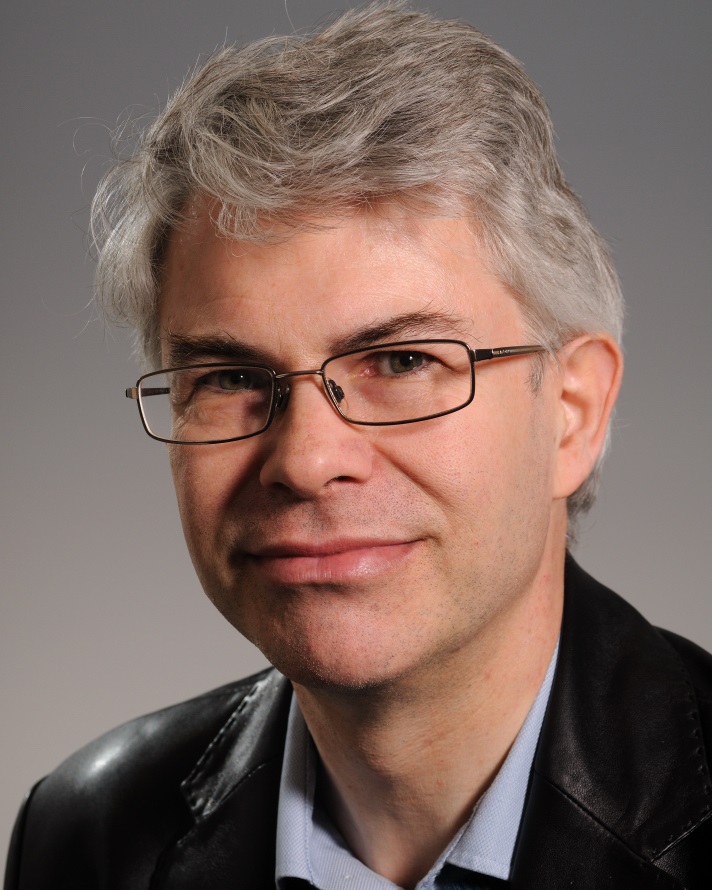}}]{Mark D. Plumbley}
(S'88-M'90-SM'12-F'15) received
the B.A.(Hons.) degree in electrical sciences
and the Ph.D. degree in neural networks from University
of Cambridge, Cambridge, U.K., in 1984
and 1991, respectively. 
Following his PhD, he became a Lecturer at King's College London, 
before moving to Queen Mary University of London in 2002. 
He subsequently became Professor and Director of the Centre for Digital Music, before joining the University of Surrey in 2015 as 
Professor of Signal Processing.
He is known for his work on analysis and processing of audio and music, 
using a wide range of signal processing techniques, 
including matrix factorization, sparse representations, and deep learning.
He is a co-editor of the recent book on 
Computational Analysis of Sound Scenes and Events,
and Co-Chair of the recent DCASE 2018 Workshop on Detection and Classifications of Acoustic Scenes and Events.
He is a Member of the 
IEEE Signal Processing Society Technical Committee on 
Signal Processing Theory and Methods, 
and a Fellow of the IET and IEEE.
\end{IEEEbiography}

\end{document}